\documentclass[preprintnumbers, floatfix, preprintnumbers, letterpaper, superscriptaddress,nofootinbib, onecolumn]{revtex4}
\pdfoutput=1
\usepackage{graphicx}
\usepackage{microtype}
\usepackage{amsmath}
\usepackage{amssymb}
\usepackage{subfigure}
\usepackage{hyperref}
\usepackage{url}
\usepackage{xcolor}
\usepackage{color}
\usepackage{mathrsfs}
\usepackage{calrsfs}
\usepackage{amsfonts}
\usepackage{latexsym}
\usepackage{ragged2e}
\usepackage{textcomp}
\usepackage{phaistos}
\usepackage{epstopdf}
\makeatletter
\renewcommand\@makefnmark{\hbox{\@textsuperscript{\normalfont\color{purple}\@thefnmark}}}
\renewcommand\@makefntext[1]{%
  \parindent 1em\noindent
            \hb@xt@1.8em{%
                \hss\@textsuperscript{\normalfont\@thefnmark}}#1}
\makeatother
\usepackage{caption}
\DeclareCaptionJustification{justified}{\leftskip=0pt \rightskip=0pt \parfillskip=0pt plus 1fil}
\definecolor{vividviolet}{rgb}{0.62, 0.0, 1.0}
\definecolor{amaranth}{rgb}{0.9, 0.17, 0.31}
\definecolor{palatinateblue}{rgb}{0.15, 0.23, 0.89}
\definecolor{brightpink}{rgb}{1.0, 0.0, 0.5}
\definecolor{cornflowerblue}{rgb}{0.39, 0.58, 0.93}
\definecolor{deepcarminepink}{rgb}{0.94, 0.19, 0.22}
\definecolor{radicalred}{rgb}{1.0, 0.21, 0.37}

\hypersetup{ linktoc=all,
    colorlinks, linkcolor={palatinateblue},
    citecolor={brightpink}, urlcolor={amaranth}
}

\graphicspath{{Images/}}

\renewcommand{\d}[1]{\ensuremath{\operatorname{d}\!{#1}}}

\graphicspath{{Images/}}
\renewcommand{\d}[1]{\ensuremath{\operatorname{d}\!{#1}}}
\makeatletter
\def\@fnsymbol#1{\ensuremath{\ifcase#1\or $\PHplaneTree$ \or $\textleaf$ 
\else\@ctrerr\fi}}%

\makeatother

\def\sideremark#1{\ifvmode\leavevmode\fi\vadjust{\vbox to0pt{\vss
 \hbox to 0pt{\hskip\hsize\hskip1em
 \vbox{\hsize1.5cm\tiny\raggedright\pretolerance10000
 \noindent #1\hfill}\hss}\vbox to8pt{\vfil}\vss}}}%
\def\sideremark#1{\ifvmode\leavevmode\fi\vadjust{\vbox to0pt{\vss
 \hbox to 0pt{\hskip\hsize\hskip1em
 \vbox{\hsize1.3cm\tiny\raggedright\pretolerance10000
 \noindent #1\hfill}\hss}\vbox to8pt{\vfil}\vss}}}%
\newcommand{\red}{\textcolor[rgb]{1.00,0.00,0.00}}

\begin{document}

\title{How Anti-de Sitter Black Holes Reach Thermal Equilibrium}

\author{Ru \surname{Ling}}
\affiliation{Center for Gravitation and Cosmology, College of Physical Science and Technology, Yangzhou University, \\180 Siwangting Road, Yangzhou City, Jiangsu Province  225002, China}

\author{Hao \surname{Xu}}
\email{haoxu@yzu.edu.cn}
\affiliation{Center for Gravitation and Cosmology, College of Physical Science and Technology, Yangzhou University, \\180 Siwangting Road, Yangzhou City, Jiangsu Province  225002, China}

\author{Yen Chin \surname{Ong}}
\email{ycong@yzu.edu.cn}
\affiliation{Center for Gravitation and Cosmology, College of Physical Science and Technology, Yangzhou University, \\180 Siwangting Road, Yangzhou City, Jiangsu Province  225002, China}
\affiliation{Shanghai Frontier Science Center for Gravitational Wave Detection, School of Aeronautics and Astronautics, Shanghai Jiao Tong University, Shanghai 200240, China}

\begin{abstract}
It is commonly known in the literature that large black holes in anti-de Sitter spacetimes (with reflective boundary condition) are in thermal equilibrium with their Hawking radiation. 
Focusing on black holes with event horizon of toroidal topology, we study a simple model to understand explicitly how this thermal equilibrium is reached under Hawking evaporation. It is shown that it is possible for a large toroidal black hole to evolve into a small (but stable) one. 
\end{abstract}

\maketitle

\section{Introduction: Thermal Equilibrium of Black Holes in Anti-De Sitter Spacetimes}

Black holes in anti-de Sitter (AdS) spacetimes have been well studied in recent decades due to their applications in holography (gauge/gravity duality). These black holes behave rather differently from their asymptotically flat counterpart. Notably, their event horizon need not be spherical, topologically speaking. Instead, black holes with hyperbolic or toroidal horizon are also valid solutions to the Einstein field equations. Regardless of their horizon topology, AdS black holes possess very different thermodynamical behavior compared to the asymptotically flat ones. The Hawking temperature in $d$-dimensions takes the form \cite{9808032} (in the units $G=c=\hbar=k_B=1$)
\begin{equation}\label{temp}
T=\frac{  k(d-3)L^2+(d-1)r_h^2 }{4\pi L^2 r_h},
\end{equation}
where $k=+1,0,-1$ correspond to horizons that are positively curved, flat, and negatively curved, respectively, and $r_h$ denotes the radial location of the event horizon. For a sufficiently large black hole, namely those with horizon size larger than the AdS curvature length scale ($r_h > L$), the temperature is directly proportional to $r_h$. That is to say, large AdS black holes are ``hot''\footnote{Even though the temperature of AdS black holes can be arbitrarily high from the viewpoint of the global geometry, local observers never see thermal radiation at such Hawking temperature \cite{0709.3738,0805.1876,0911.4144}.  Keeping this subtlety in mind, we shall no longer put scare quotes around the words hot or cold hereinafter.}. This lies in stark contrast with the asymptotically flat Schwarzschild black hole, whose temperature scales inversely proportional to its mass (and therefore size). 

In addition, asymptotically locally AdS spacetimes have a timelike boundary at spatial infinity. Remarkably, null geodesics from within the bulk can hit the boundary and be reflected back in a finite affine parameter interval (and also in a finite coordinate time $t$, if we use the canonical Schwarzschild-like coordinates). To see this, let us focus on the $k=0$ case, which is widely used in holography. 
Hereinafter, we shall refer to such black holes as ``flat black holes''.

Suppose there is no black hole. The metric tensor
\begin{equation}
\d s^2 = -\frac{r^2}{L^2}\d t^2 + \frac{L^2}{r^2}\d r^2 + r^2 \left(\sum_{i=1}^{d-2} (\d x^i)^2\right),
\end{equation} 
simply describes a flat foliation of the maximally symmetric AdS spacetime.
This coordinate system fails at the center $r=0$, so let us consider $r = \varepsilon>0$, where $\varepsilon$ is small.
The proper time between any two events both located at $r = \varepsilon$ is $(\varepsilon/L)\Delta t$, where $\Delta t$ for a photon that goes from $r=\varepsilon$ to $\infty$ and back is
\begin{equation}
\Delta t=2\int_\varepsilon^\infty \frac{L^2}{r^2} \d r = \frac{2L^2}{\varepsilon}.
\end{equation}
This is finite, although large.
The proper time elapsed for the static observer is $2L$. Note that $\varepsilon$ drops out in the proper time, as it should, since the AdS ``center'' is arbitrary. 
As a consequence, if a reflective boundary condition is imposed, the Hawking photons will be reflected back into the black hole and so a sufficiently large black hole can attain thermal equilibrium. (Another consequence is the non-linear instability of AdS: surprisingly a large class of arbitrarily small perturbations can be reflected and refocused in the bulk, thus causing black hole formation \cite{1104.3702}. See also the review \cite{1708.05600}.)

For the $k=1$ case, small black holes are also hot ($T\sim 1/r_h$ as can be seen from Eq.(\ref{temp})), much like a small asymptotically flat Schwarzschild black hole. Since Hawking radiation takes time to hit the boundary and be reflected back, such small black holes can therefore completely evaporate before they have any hope to achieve thermal equilibrium. 
In other words, large black holes (which have positive specific heat) are stable while small black holes (which have negative specific heat) are therefore unstable. We could in principle use this stability criterion to define ``large'' and ``small''. While this criterion happens to coincide with using either the mass or the horizon size being greater than $L$ to define the black hole ``size'' in the $k=1$ case, it does not hold for the $k=0$ case that we would like to focus on in this work. These black holes have Hawking temperature that is proportional to $r_h$ regardless of the black hole size, this means that small flat black holes are cold, i.e., their rate of evaporation is slow. Therefore it is not impossible for small black holes to attain thermal equilibrium with their Hawking radiation. \emph{All} $k=0$ black holes would therefore be ``large'' if we were to use the stability/specific heat to define its ``size''. This is why we use $r_h > L$ as the definition of a large black hole.

Furthermore, since the boundary condition can be changed to a completely absorptive one in which there is no thermal equilibrium (see Sec.(\ref{1})), we prefer to use a definition that holds independent of the boundary condition. That is, a large black hole would remain large even if we change the boundary condition. Given a fixed black hole in the bulk, it takes time for the radiation to reach the boundary and come back. Until the radiation reaches the boundary (and potentially reflected back or absorbed depending on the boundary condition), the black hole has no knowledge of whether it can reach equilibrium. So a local criteria that allows us to define the black hole size at any given time, even \emph{before} the first Hawking radiation is emitted (so that we can speak of whether an \emph{initially} ``large'' black hole can evolve into a ``small'' one or remains ``large'', even in the $k=1$ case) is more useful. As we shall see, defined this way, large flat black holes can evaporate into a small black hole which is in thermal equilibrium with their Hawking quanta\footnote{Gibbons and Perry argued that black holes can remain in thermal equilibrium with a heat bath even in the presence of particle interactions, though his work is restricted to the asymptotically flat case, the conclusion is likely to be generic \cite{3343}.}.

Indeed, our objective in this work is to explicitly study a simplified model that allows us to see how AdS black holes attain thermal equilibrium. The end result is perhaps expected, but the detailed evolution -- to our knowledge -- has not been explicitly studied.

\section{Evaporation Under A Completely Absorptive Boundary Condition}\label{1}

For comparison purpose, it is instructive to first review the evaporation of flat black holes under the assumption that the boundary is completely \emph{absorptive}, which was investigated in details in \cite{1507.07845}. 
In holography this can be achieved by coupling the boundary with another auxiliary CFT \cite{1304.6483,0804.0055,1307.1796}. We remark that the boundary condition of AdS is not only important classically (since AdS spacetime is not globally hyperbolic), but also crucial for a consistent quantization scheme of the fields in the bulk \cite{PhysRevD.18.3565}. Although boundary conditions are mathematically arbitrary, physical considerations would dictate which type of conditions should be chosen. For example, an absorptive boundary condition is imposed when one wants to allow large black holes to evaporate (to study Hawking evaporation and their associated phenomena). In the context of the study of perturbations in the AdS bulk, the situation is a lot more complicated (to ensure the perturbations have the correct asymptotic behavior) -- see, for example, the highly nontrivial work of Ishibashi and Wald, which determined all possible boundary conditions that can be imposed at infinity for scalar, electromagnetic, and gravitational perturbations of AdS spacetime \cite{0402184}; see also \cite{1302.1580} for Kerr-AdS perturbations. Boundary conditions have also been studied from the perspective of the Hamiltonian approach \cite{1701.01119}.  Here, our concern is somewhat simpler: we are only interested in the mass evolution of the black hole $M(t)$ under Hawking evaporation, given a specified boundary condition.

Let us focus on the 4-dimensional case, in which a static black hole of mass $M$ with toroidal topology 
 is described by the metric tensor \cite{9404041,Huang:1995zb,9609065}
\begin{equation}\label{metric}
\d s^{2}=-\left(\frac{r^{2}}{L^{2}}-\frac{2 M}{\pi K^{2} r}\right) \d t^{2}+\left(\frac{r^{2}}{L^{2}}-\frac{2 M}{\pi K^{2} r}\right)^{-1} \d r^{2}+r^{2}\left(\d x^{2}+\d y^{2}\right),
\end{equation}
where $K$ is the compactification parameter: if we consider the horizon to be a square flat torus $T^2 = S^1 \times S^1$, then each $S^1$ has  
circumference 2$\pi K$. The coordinates $x,y$ are the usual coordinates on a (compactified) plane. One can of course consider taking the planar limit to obtain a black brane.

The black hole radius $r_h$ and the Hawking temperature are given by
\begin{equation}
r_{h}=\left(\frac{2 M L^{2}}{\pi K^{2}}\right)^{\frac{1}{3}}\label{rhh}, \quad T=\frac{3 r_{h}}{4 \pi L^{2}}=\frac{3 M}{2 \pi^{2} K^{2} r_{h}^{2}}.
\end{equation}
The effective potential for massless particles in the background of a toroidal black hole geometry does not have a local maximum \cite{1403.4886,9803061}. It is explicitly given by
\begin{equation}
V(r)=\frac{J^{2}}{r^{2}}\left(\frac{r^{2}}{L^{2}}-\frac{2 M}{\pi K^{2} r}\right),
\end{equation}
where $J$ is the angular momentum of the particle. This expression increases monotonically and asymptotically tends to a constant value $J^2/L^2$ near the boundary. The relevant area $\sigma$ that one should use in the Stefan-Boltzmann law for luminosity, $\sigma T^4$, for the Hawking emission is $4\pi^2K^2 L^2$, which is fixed by the cosmological constant instead of the black hole mass \cite{1403.4886, 9803061}. To understand this, we note that a Hawking particle has to overcome a potential with height $J^2/L^2$ in order to reach null infinity. In other words, a particle must have a ratio of angular momentum to energy of at least $L$ to escape. The ``cross section'' defined with this scale is therefore proportional to\footnote{The derivation and argument in \cite{1403.4886} is incorrect, though the result is correct, and the general idea is contained therein.} $L^2$, which agrees with the result obtained by the rigorous quantum approach of \cite{9803061}. 

Applying the Stefan-Boltzmann law, the mass loss equation is 
\begin{equation}
\frac{\d M}{\d t}=-a \alpha \sigma T^{4},\label{dmdt}
\end{equation}
where $a$ is the radiation constant and $\sigma=4\pi^2 K^2 L^2$ is the cross section defined above. The greybody factor $\alpha$ depends on the type of emitted particles, but if the greybody factor is not equal to 1, it will have an effect on the black hole to prolong the black hole lifetime. Since we are only interested in the qualitative picture of the evolution, let us thus ignore the prefactors and simplify the equation as 

\begin{equation}
\frac{\d M}{\d t} \sim-\frac{L^{2} M(t)^{4}}{r_{h}^{8}(t)}=-\frac{M(t)^{\frac{4}{3}}}{L^{\frac{10}{3}}}.
\end{equation}
Integrating the above formula we see that the black hole lifetime is infinite, since
\begin{equation}
M(t)=\left[\frac{3}{L^{-10/3}t+3M_0^{-1/3}}\right]^3,
\end{equation}
(where $M_0$ is the initial mass) tends to zero only when $t \to \infty$.

\begin{figure}[thbp]
\center{
  \includegraphics[scale=0.6]{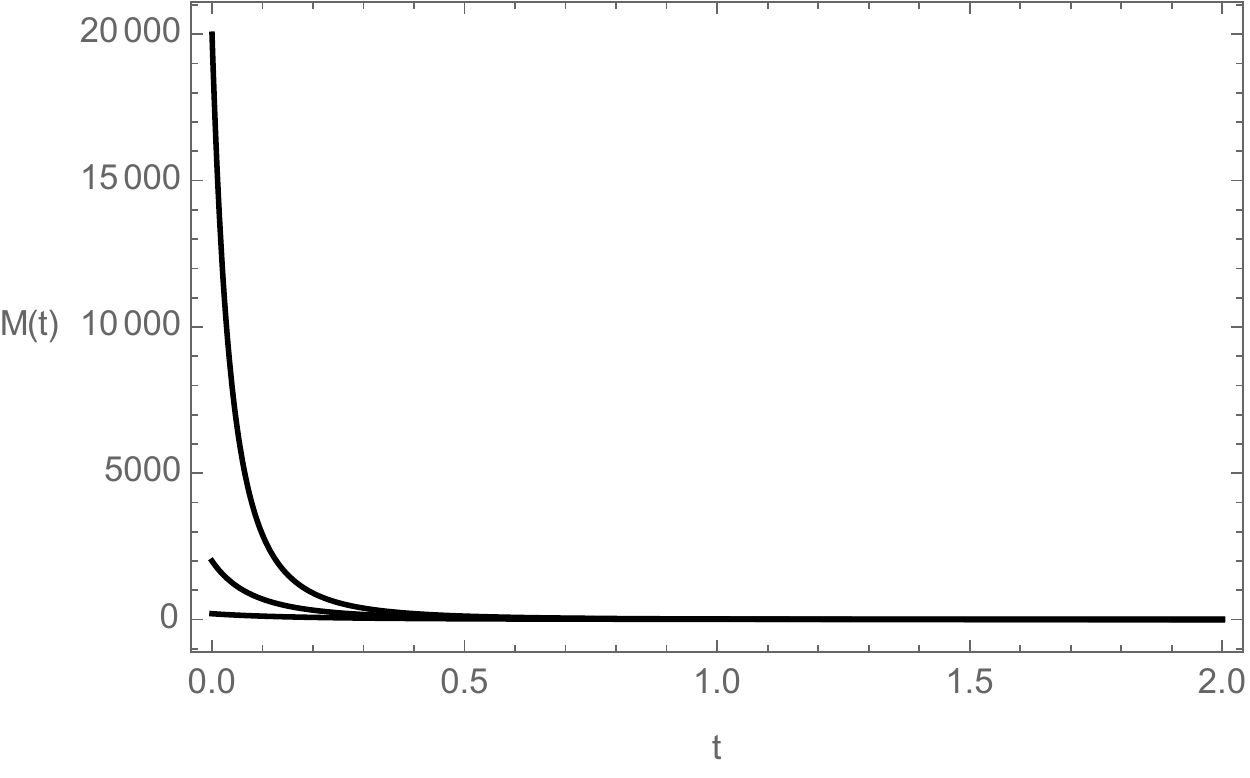}
  \caption{The evolution of toroidal AdS black hole under the completely absorptive boundary condition. From top to bottom the initial masses are chosen to be 20000, 2000, 200 respectively.}\label{figurea}}
\end{figure}

In Fig. (\ref{figurea}) we present some examples of the evolution of the toroidal black holes with different initial masses under the completely absorptive boundary condition. Hereinafter, we set $L=1$ for further simplification. Note that arbitrarily large black holes can lose a lot of mass initially due to their high temperature. In fact, regardless of their initial mass, all toroidal black holes take about the same time of the order of $L^3$ to evaporate down to $M \sim L$ \cite{1507.07845}. (For the $k=1$ case arbitrarily large black holes evaporate in a finite time of order $L^3$, as shown by Page \cite{1507.02682}.) When the black hole becomes smaller, the temperature also decreases, so
the evaporation process is increasingly difficult. This obeys the third law of black hole thermodynamics -- the zero temperature state, which corresponds to a zero mass black hole, is unattainable in a finite time.

\section{Thermal Equilibrium Under A Completely Reflective Boundary Condition}

We now consider the case in which the boundary is completely reflective, as is usually assumed in holography. 
We shall model the evaporation process with a delayed differential equation (DDE):
\begin{equation}\label{DDE}
\frac{\d M(t)}{\d t}=\left\{\begin{array}{lc}
-\frac{M(t)^{\frac{4}{3}}}{L^{\frac{10}{3}}}, &   t<t^* ,\\
-\frac{M(t)^{\frac{4}{3}}}{L^{\frac{10}{3}}}+\frac{M\left(t-t^{*}\right)^{\frac{4}{3}}}{L^{\frac{10}{3}}}, &   t>t^*,
\end{array}\right.
\end{equation}
where $t^*$ denotes the time it took for an emitted particle to reach the boundary and be reflected back into the black hole horizon. 
In other words, $t^*$ is the round trip duration, analogous to $\Delta t$ in the empty AdS spacetime we computed in the Introduction. 

The DDE describes the following process: initially when the black hole starts to emit Hawking radiation (say just after its formation), Hawking radiation takes time to travel to the boundary and back. Therefore, prior to time $t^*$, the black hole does not absorb any incoming radiation. After time $t>t^*$, it continues to emit at the temperature at time $t$. The radiation that returns from infinity, however, has a higher temperature since it was emitted at an earlier time $t-t_*$. The redshift of the outgoing photon is canceled by the blueshift on its return trip. To be precise, the cancellation is only perfect if we consider a static black hole. Here the black hole radius would have changed after time $t^*$, and thus the blueshift would not perfectly compensate for the redshift (see more in the Discussion for other subtleties). Still, it is insightful to first study the most simplified model that captures the essence of the problem.

We will consider two cases: fixed $t^*$ and varying $t^*$. Indeed, as the black hole evaporates, its size changes, and $t^*$ will change accordingly. Thus, the varying $t^*$ case is more physical. However, as we shall see, even fixing $t^*$ for simplicity still captures some of the important physical features. This suggests that despite the simplification of the model it is somewhat robust.

\subsection{Case I: Fixed $t^*$}

\begin{figure}[thbp]
\center{
  \includegraphics[scale=0.38]{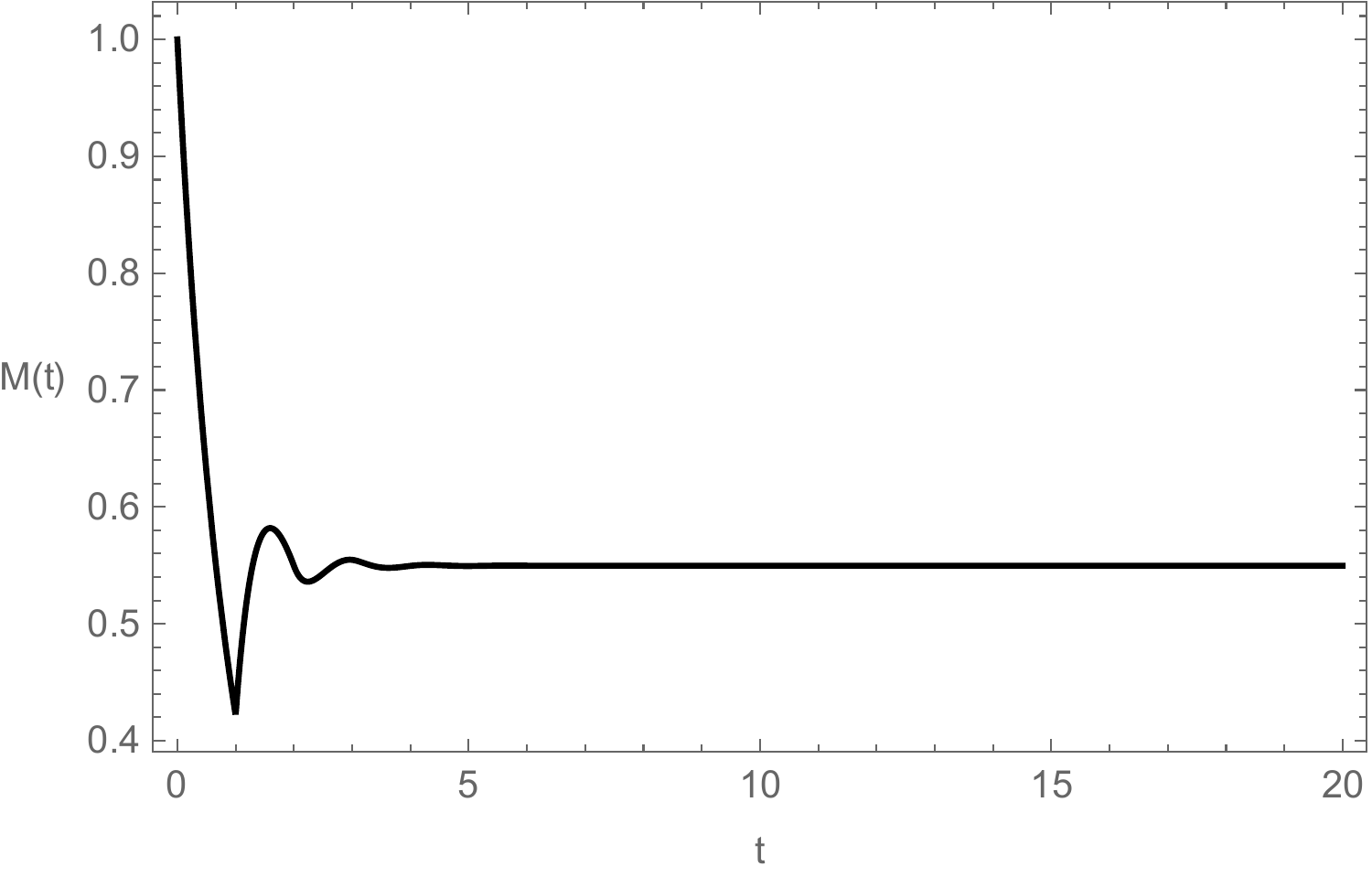}
  \includegraphics[scale=0.38]{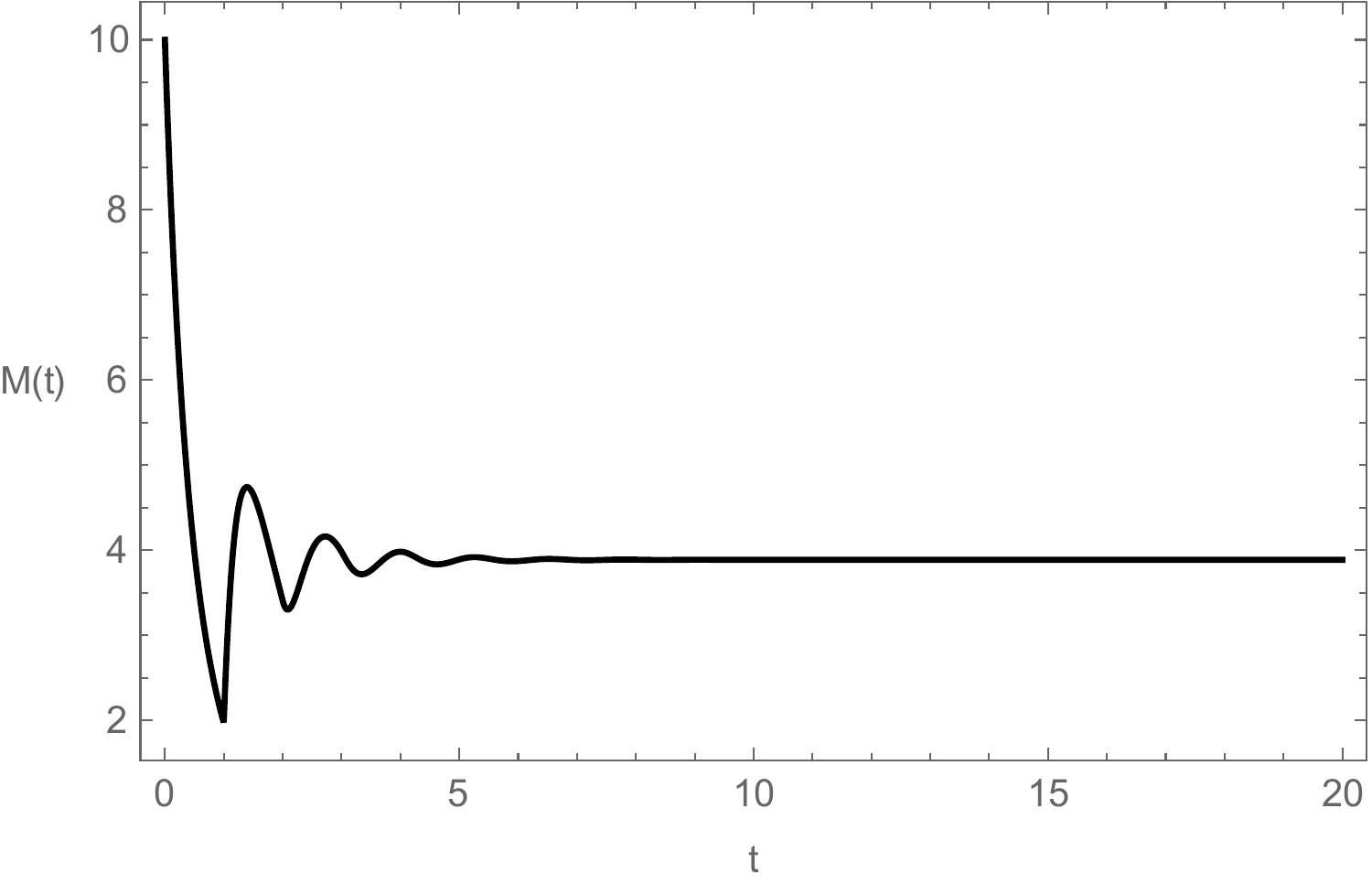}
  \includegraphics[scale=0.38]{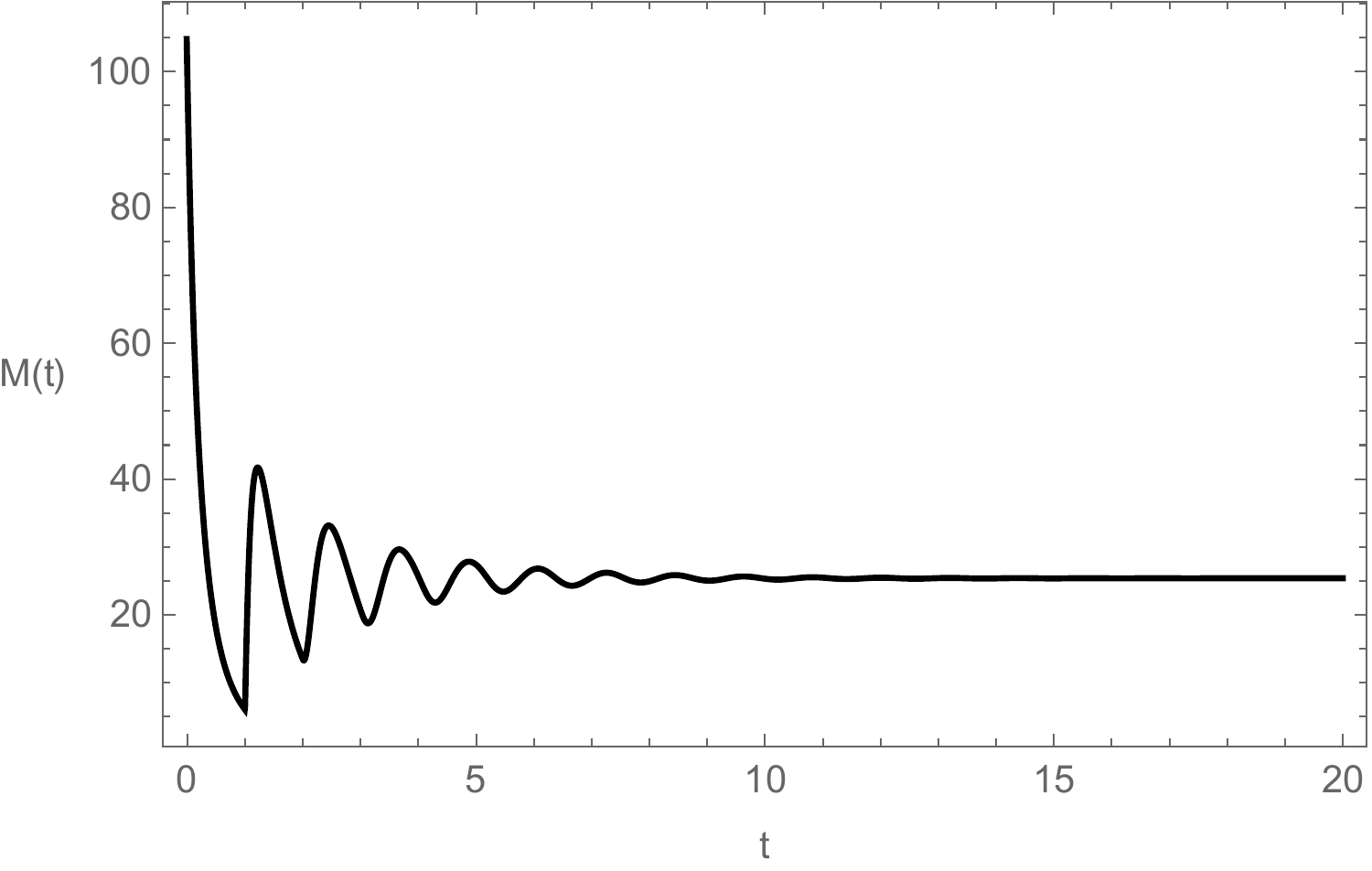}
  \caption{The black hole evolution at $t^*=1$. From left to right the initial mass of black hole are $M_\text{ini}=1,10,100$ respectively.}\label{figureh}}
\end{figure}

\begin{figure}[thbp]
\center{
  \includegraphics[scale=0.38]{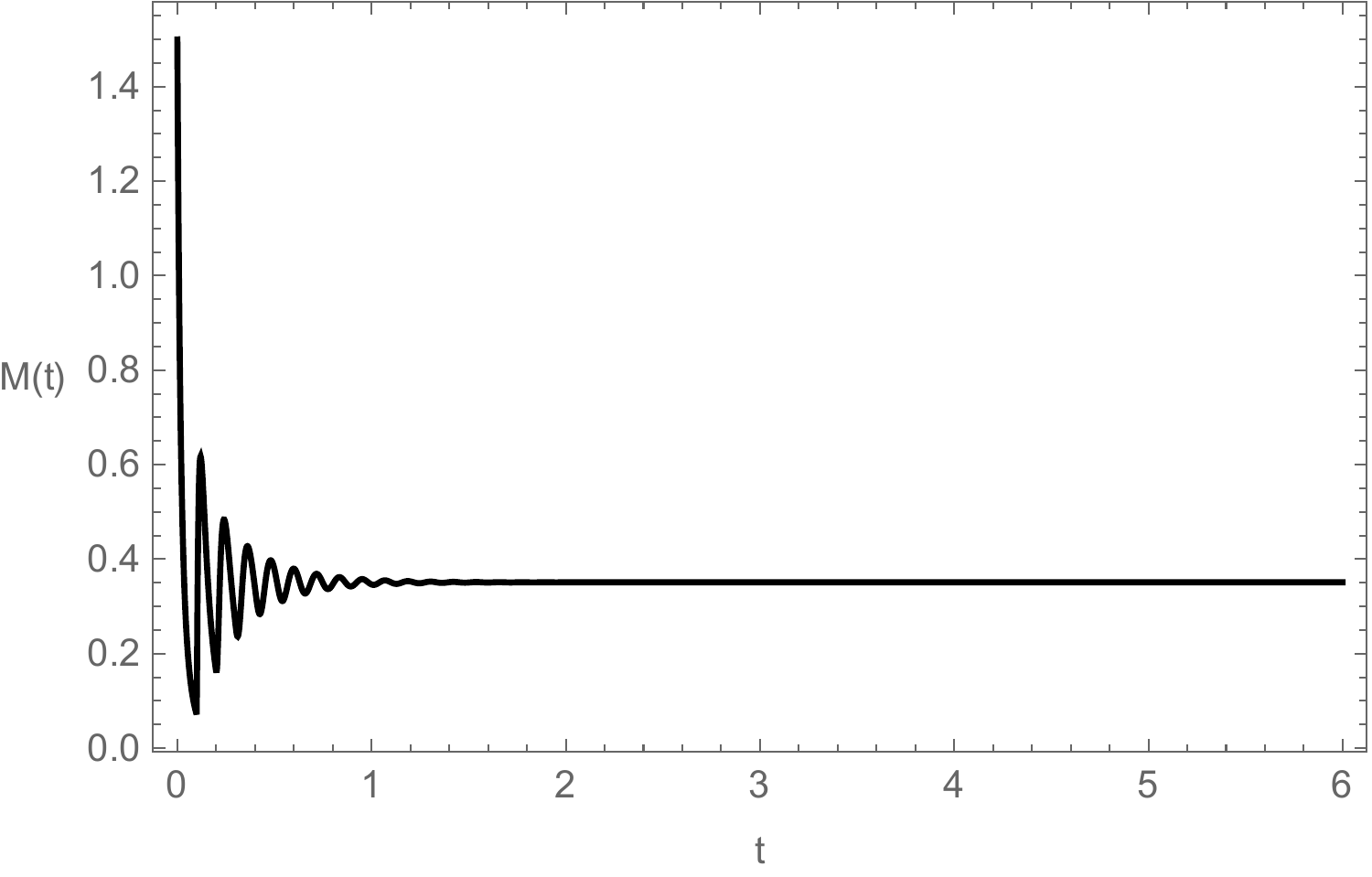}
  \includegraphics[scale=0.38]{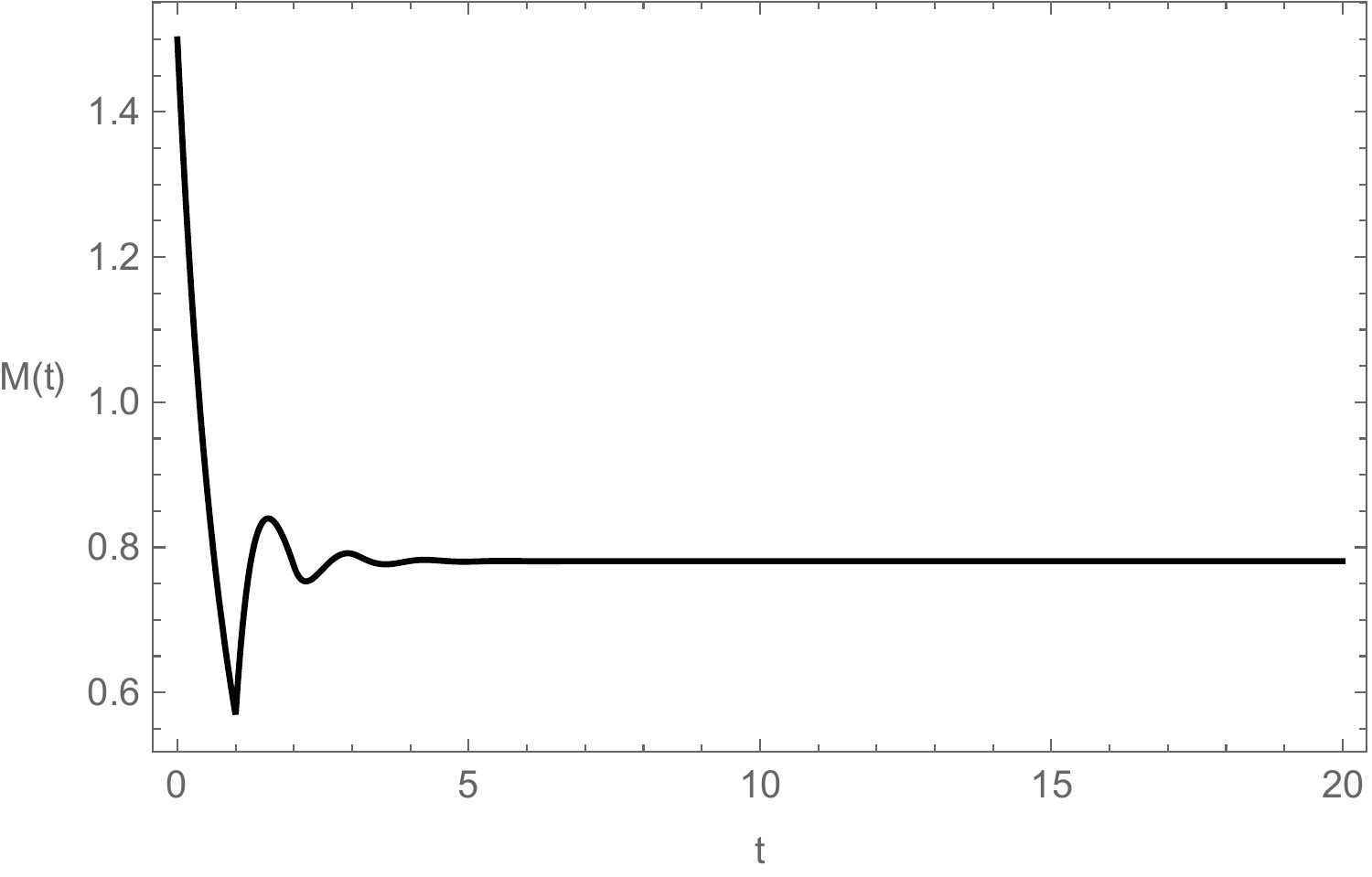}
  \includegraphics[scale=0.38]{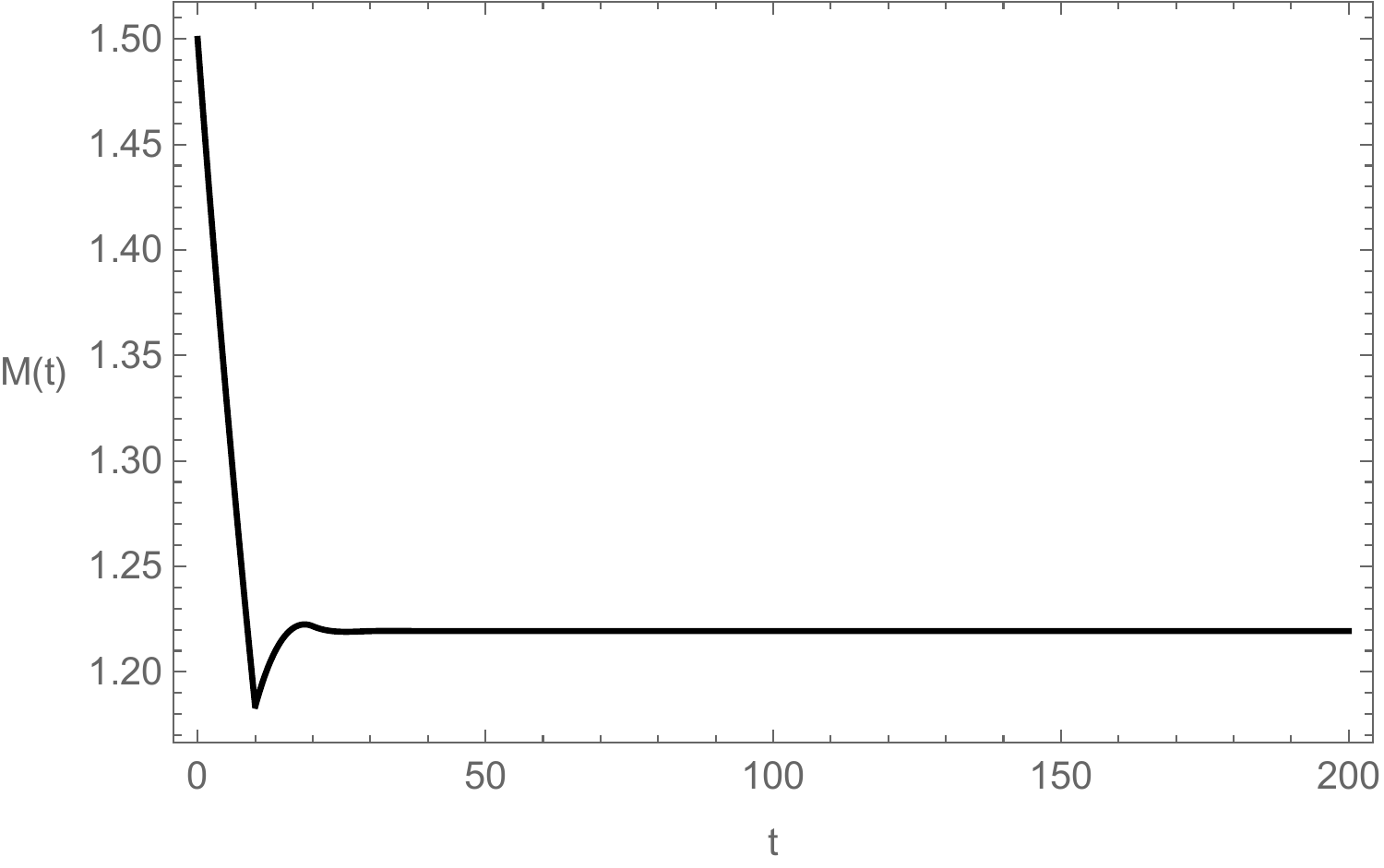}
  \caption{
The black hole evolution at $M_\text{ini}=1.5$. From left to right the $t^*= 0.1,1,10$ respectively.}\label{figurehh}}
\end{figure}

Let us fix $t^*$ to be a constant and solve the DDE numerically with MAPLE. In Fig. (\ref{figureh}) and Fig. (\ref{figurehh}) we present some examples of the black hole evolution for various values of $t^*$ and $M_\text{ini}$. We see that the black hole mass fluctuates with time, which is very similar to the echo behavior found in \cite{Saraswat:2019npa} (see their Fig. (6)). In Fig. (\ref{figureh}), as the initial mass of the black hole increases, the oscillations persist for a longer duration, which means that it is harder for the black hole to reach a stable equilibrium state. On the other hand, in Fig. (\ref{figurehh}), as $t^{*}$ increases, we observe that the oscillation frequency decreases. In other words, the longer it takes the Hawking particle to complete its round trip, the less oscillation is observed in the mass evolution.

In these plots, we observe that $t^*$ corresponds to the first ``bounce'' in the mass function $M(t)$. If $t_* \to \infty$, the curve would have continued to decrease to zero asymptotically, thus recovering the result for completely absorptive boundary condition we reviewed in Sec.(\ref{1}). We note from the plots that this bounce is not smooth. The physical reason is this: a toroidal black hole has temperature directly proportional to its mass, so before $t^*$, the emitted Hawking quanta are relatively hot. By the time the black hole re-absorbed these particles at $t=t^*$, there is a huge jump in the energy (mass) of the black hole as the hot radiation suddenly dumped into the horizon. Depending on the mass and the temperature of the black hole, the subsequent bounces can still fail to be smooth, though eventually as the mass fluctuation gets smaller, the bounce will become smoother.

In the literature, sometimes one defines ``large'' black hole by the condition $M > L$ and ``small'' black hole by $M < L$. 
With this criterion, in the middle diagram of Fig. (\ref{figurehh}), we have an example of a large black hole equilibrating into a small black hole under Hawking evaporation. 
Since $r_h$ is related to $M$ by Eq. ($\ref{rhh}$), $M > L$ also means $r_h > (2/(\pi K^2))^{1/3}L \gtrsim L$ (for $K=1$), so it does not matter if we use $r_h$ or $M$ to characterize ``large'' and ``small''.
It is true that for larger values of $K$, these two characterizations could differ, but the point is that no matter which variable we use in the evolution, $M$ or $r_h$, it is always possible to choose a suitable initial value such that transition from a large black hole to a small one is possible.

Note that oscillations in the mass never truly dies off completely, so we need a practical definition of what thermal equilibrium means. Denote the set of local maxima by $\left\{M^\text{max}_i\right\}$ and the set of local minima by $\left\{M^\text{min}_i\right\}$, $i \in \mathbb{N}$. Let us denote the time that corresponds to either a local maximum or a local minimum by $t_i$.
We will consider a black hole to have reached thermal equilibrium with a stable mass $M_\text{stb}$, if there exists $N\in \mathbb{N}$ such that $\left(M^\text{max}_i-M^\text{min}_j\right)/M_\text{stb}<0.001\%$ for all $i, j \geqslant N$. We refer to $t_N$ as the stable time, henceforth denoted by $t_\text{stb}$, at which the mass is  none other than $M_\text{stb}$. That is, we say that equilibrium/stable state has been reached when the ratio of the mass fluctuation to the stable mass is sufficiently small (this cutoff can be made arbitrarily small if one wishes, but in practice one should work within the numerical accuracy/resolution).

\begin{figure}[thbp]
\center{
  \includegraphics[scale=0.4]{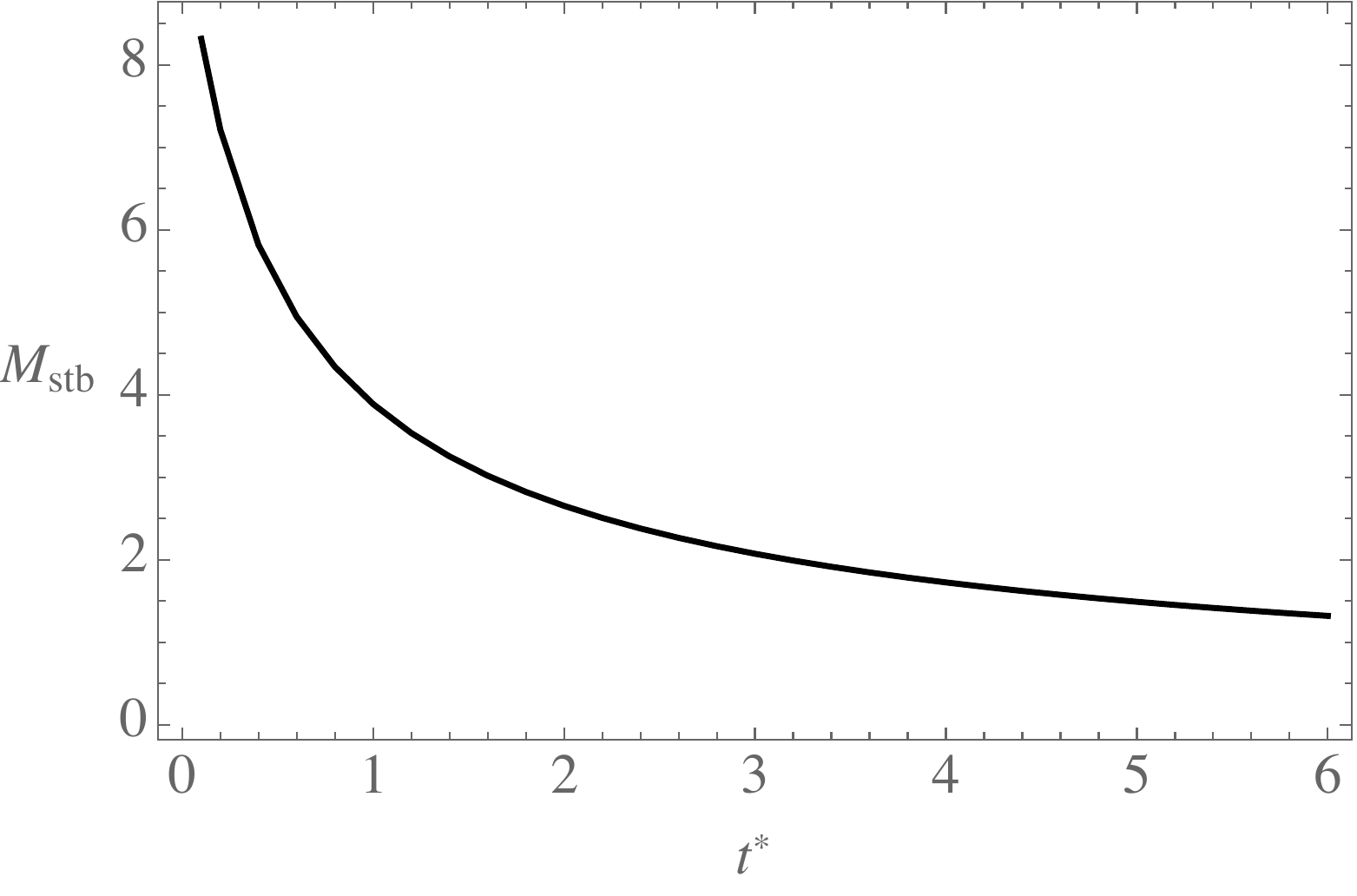}
  \includegraphics[scale=0.41]{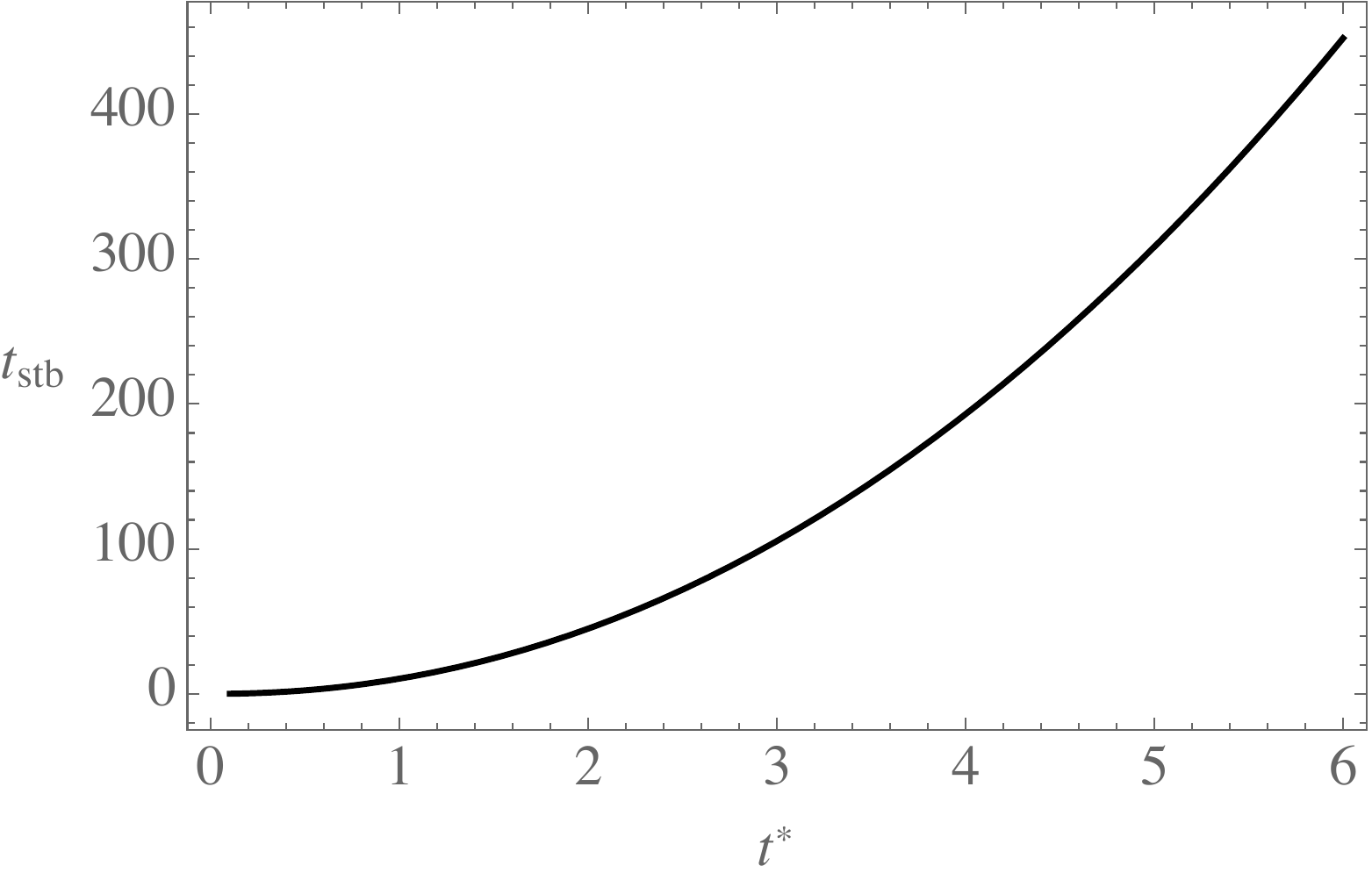}
  \caption{\textbf{Left:} The relationship between $t^{*}$ and the stable mass $M_\text{stb}$. \textbf{Right:} The relationship between $t^{*}$ and the stable time $t_\text{stb}$. In both figures we set $M_\text{ini}=10$ and $L=1$.}\label{figuredd}}
\end{figure}

In Fig. (\ref{figuredd}), we studied the behaviors of the final stable mass $M_\text{stb}$ and the corresponding thermal equilibrium time $t_\text{stb}$ as we change the values of the fixed $t^{*}$. As can be seen from the left figure, for a larger value of $t^*$ we always have a smaller $M_\text{stb}$. This is because as $t^*$ gets larger, more particles are dissipated  in the process of Hawking emission, and less mass is re-absorbed by the black hole. In the right figure, we observe that $t_\text{stb}$ becomes larger when $t^*$ increases. Combining with Fig. (\ref{figurehh}), we see that a smaller $t^*$ implies a higher frequency of mass oscillation. This is to be expected since a smaller $t^*$ makes it easier for the radiating particles to be reflected back and thus re-absorbed by the black hole. The radiation ``bouncing'' back and forth between the horizon and the boundary thus causes the rapid mass oscillation. On the contrary, a larger $t^*$ implies a longer journey time for the emitted particle, and the oscillation frequency is also smaller, thus corresponding to a longer time for reaching thermal equilibrium.

\begin{figure}[thbp]
\center{
  \includegraphics[scale=0.433]{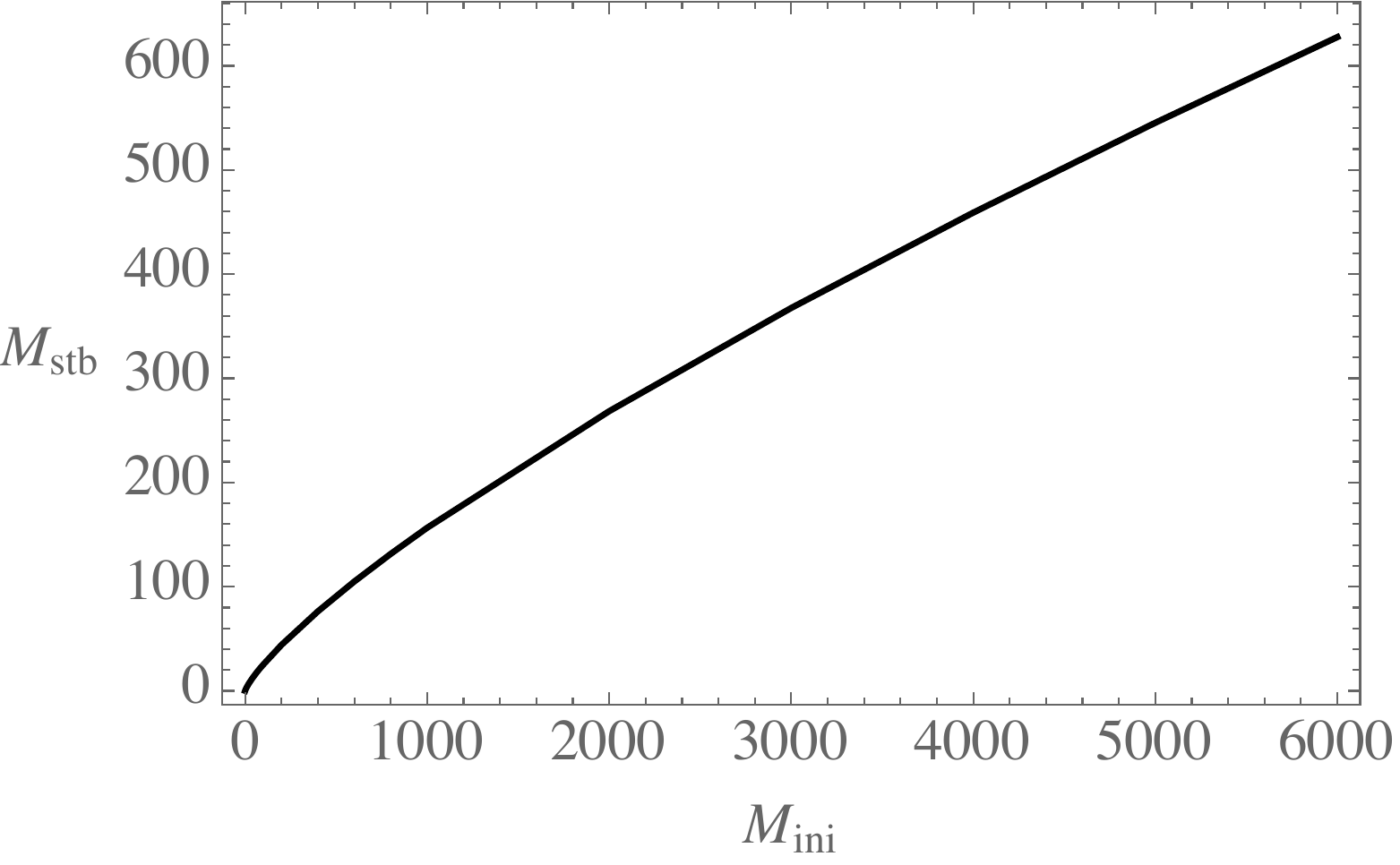}
  \includegraphics[scale=0.421]{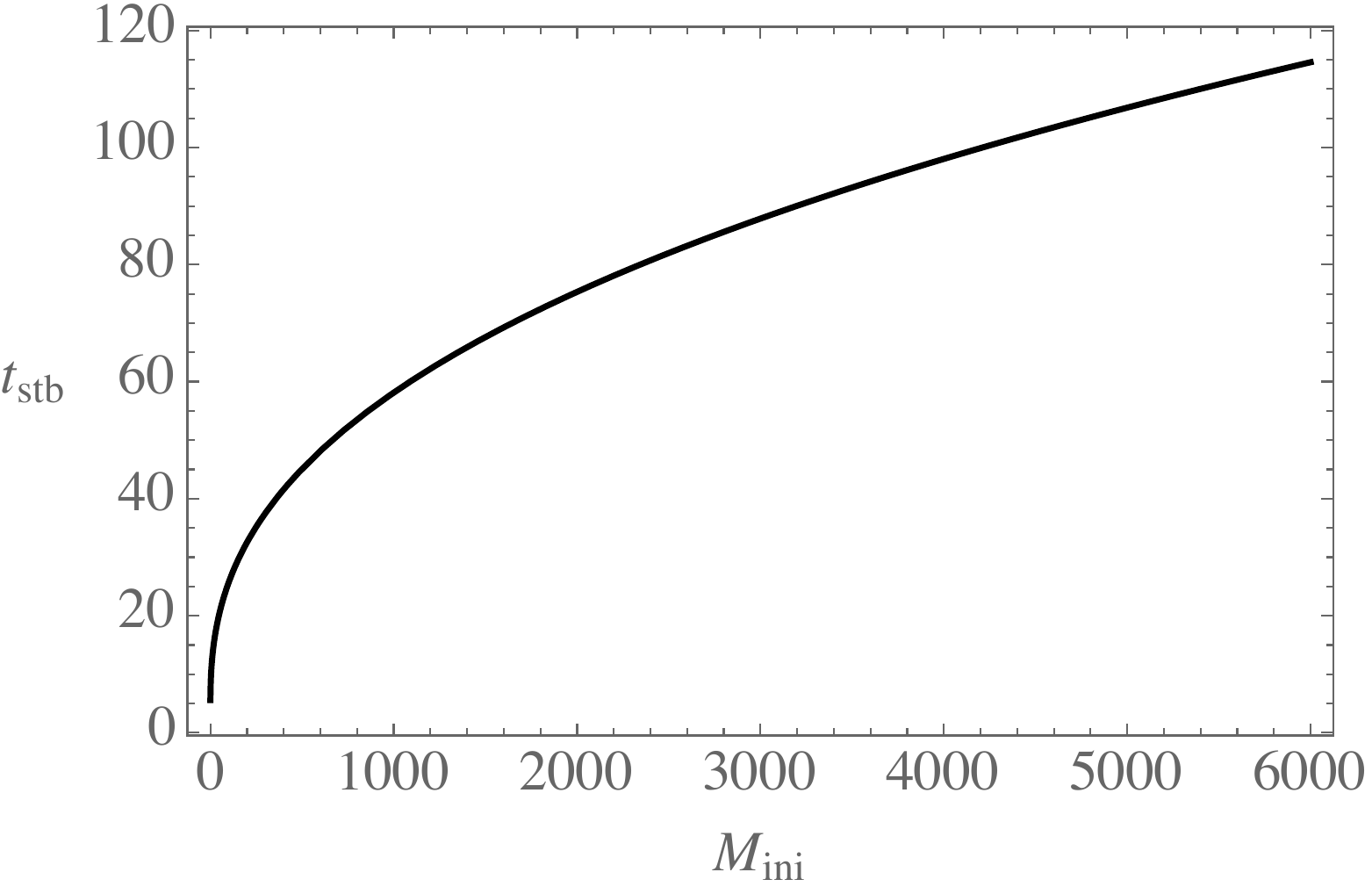}
  \caption{Left: The relationship between the initial mass $M_\text{ini}$ and final stable mass $M_\text{stb}$; Right: The relationship between the initial mass $M_\text{ini}$ and the final stable time $t_\text{stb}$. In both figures we set $t^{*}=1$ and $L=1$.}\label{figuremm}}
\end{figure}

In Fig. (\ref{figuremm}), we present the behaviors of $M_\text{stb}$ and $t_\text{stb}$ with various initial black hole mass $M_\text{ini}$ and fixed $t^*$. In the left figure, we see that the final stable mass increases with the initial mass. In the right figure, $t_\text{stb}$ is also increasing with $M_\text{ini}$. Ultimately the assumption of fixed $t^*$ is expected to break down and the results misleading when $M_\text{ini}$ grows too large. This is because -- colloquially speaking -- a very large black hole is ``closer'' to the boundary and so one expects that the time a massless particle takes to reach the infinity and back to be shorter than a smaller black hole. Hence, let us now move on to a more realistic model: allowing $t^*$ to vary as the mass changes.

\subsection{Case II: Varying $t^*$}

During the evolution of the black hole, the radius of the black hole will change due to the absorption and emission of particles. For black holes with a different radius,  the emitted particles will  take different duration to reach the AdS boundary and be reflected back. Thus the case of varying $t^*$ is more physical. Explicitly we can calculate the round trip time as $t^*$ as \cite{Saraswat:2019npa}

\begin{equation}
t^*=\int_{r_{h}+ \delta}^{\infty} \frac{2}{\frac{r^2}{L^2}-\frac{2 M}{r}} \d r=F(r)|^{\infty}_{r_h+\delta},
\end{equation}
where $\delta >0$ is a regulator to avoid divergence of the integral. This can be understood as the particles get emitted and re-absorbed at $r_h+\delta$, which is near the horizon. 
The divergence in the integral for $\delta=0$ is easy to understand: photon emitted \emph{exactly} on the horizon is not emitted outward. (Indeed, for AdS black holes, it is expected that most Hawking radiation comes from the near-horizon region  \cite{2003.10429}, the ``quantum atmosphere'' is thin \cite{1511.08221}, which is not true in general in the asymptotically flat case \cite{1511.08221, 1607.02510, 1701.06161}. In this work we nevertheless consider $\delta$ to be arbitrary, within $O(1)$ of the horizon scale, just to explore the effect on $t^*$.) Strictly speaking, this expression of $t^*$ is only true if $M$ is constant. When the mass -- and therefore the size of the black hole -- changes, one should properly calculate $t^*$ by integrating from the horizon to infinity, and then integrate from infinity back to the new position of the horizon. In our simple model this is not taken into account, which is consistent with our DDE set up that ignores the redshift and blueshift of the radiation (which only perfectly compensate each other in the static case). We hope that even with such a short-coming, our model still captures the essential main features of the physics (we can be a little more explicit; see below).

Substituting in the black hole radius $r_h=(2M L^2)^\frac{1}{3}$, the above integral $F(r)$ can be obtained as

\begin{eqnarray}\label{ff}
	F(r)&=& \int\frac{2}{\frac{r^2}{L^2}-\frac{2 M}{r}} \d r  \nonumber \\  
      &=&L^{\frac{4}{3}}\bigg(\frac{2^\frac{2}{3} \ln \left(r-2^\frac{1}{3}M^\frac{1}{3}L^{\frac{2}{3}}\right)}{3 M^ \frac{1}{3}}-\frac{2^\frac{2}{3} \ln\left(r^2+2^\frac{1}{3} M^\frac{1}{3}L^{\frac{2}{3}} r+2^\frac{2}{3} M^\frac{2}{3}L^{\frac{4}{3}}\right)}{6 M^ \frac{1}{3}}+\frac{2^\frac{2}{3} \arctan\left[\frac{1}{3}\sqrt{3} \left(\frac{2^\frac{2}{3} r}{M^ \frac{1}{3}L^{\frac{2}{3}}}+1\right)\right]}{\sqrt{3}M^ \frac{1}{3}}\bigg).
\end{eqnarray}

The function $F(r)$ tends to ${ 2^{\frac{2}{3}} \sqrt{3}L^{4/3} \pi}/6 M^{\frac{1}{3}}$ as $r\to \infty$. On the other hand, taking the series expansion at $r=r_h +\delta$ for small $\delta$ we have
\begin{equation}
F(r_h+\delta)=\frac{2^{2/3} \ln (\delta)L^{\frac{4}{3}}}{3M^{1/3}}- \frac{2^{2/3} \ln \left(3\cdot2^{2/3} M^{2/3}L^{\frac{4}{3}}\right)L^{\frac{4}{3}}}{6M^{1/3}}+\frac{2^{2/3} \pi L^{\frac{4}{3}}}{3\sqrt{3}M^{1/3}}-\frac{1}{18 M} \delta^{2}+O\left(\delta^{3}\right).
\end{equation}
Thus we can obtain the expression of $t^{*}$ in the terms of $M$ and $\delta$:
\begin{equation}
t^{*}=\frac{1}{18} \frac{2^{2/3}\pi\sqrt{3}L^{\frac{4}{3}}}{M^{1/3}}-\frac{2^{2/3} \ln (\delta)L^{\frac{4}{3}}}{3M^{1/3}}+\frac{2^{\frac{2}{3}} \ln \left(3\cdot2^{\frac{2}{3}} M^{\frac{2}{3}}L^{\frac{4}{3}}\right)L^{\frac{4}{3}}}{6M^{1/3}}.\label{tstar}
\end{equation}
Note that this is logarithmically divergent in the limit $\delta \to 0$.

From Eq. (\ref{tstar}) we should at least require that $t^*$ is positive, which yields the inequality for which the approximation is applicable
\begin{equation}
M > 0.0063\times \frac{\delta^3}{L^2}.
\end{equation}
This always holds for $\delta \sim O(1) r_h$ that we assumed. On the other hand, if the periodicity of the torus (more precisely the circumference of one of its $S^1$ direction) is comparable to or shorter than the local thermal wavelength of the Hawking radiation (i.e. the Tolman temperature, $T_\text{local} = T/\sqrt{|g_{tt}|}$), then one would expect that the model needs to take into account the discreteness of the modes \cite{1507.07845}, and our model would require corrections. This consideration can be implemented by first defining $\mu := KM$ and $\mathfrak{t}:=t/K, ~\mathfrak{r}:=Kr$, so that 
\begin{equation}
g_{\mathfrak{t}\mathfrak{t}}=-\left(\frac{\mathfrak{r}^2}{L^2} - \frac{2 \mu}{\pi \mathfrak{r}}\right).
\end{equation}
The horizon is now at $\mathfrak{r} = (2 \mu L^2/\pi)^{1/3}$, so the proper circumference of the horizon is $2 \pi (2\mu L^2/\pi)^{1/3}$.
At large $\mathfrak{r}$, the local Tolman temperature is $T_\text{local} \sim 1/L$, so that the number of thermal wavelengths within a circumference of the torus is at the order $( \mu L^2)^{1/3}/L$. Requiring that this is larger than unity to avoid the discreteness issue therefore yields $\mu > L$, or for $K=1$, the inequality $M >L$, for which our model is applicable.

\begin{figure}[thbp]
\center{
  \includegraphics[scale=0.45]{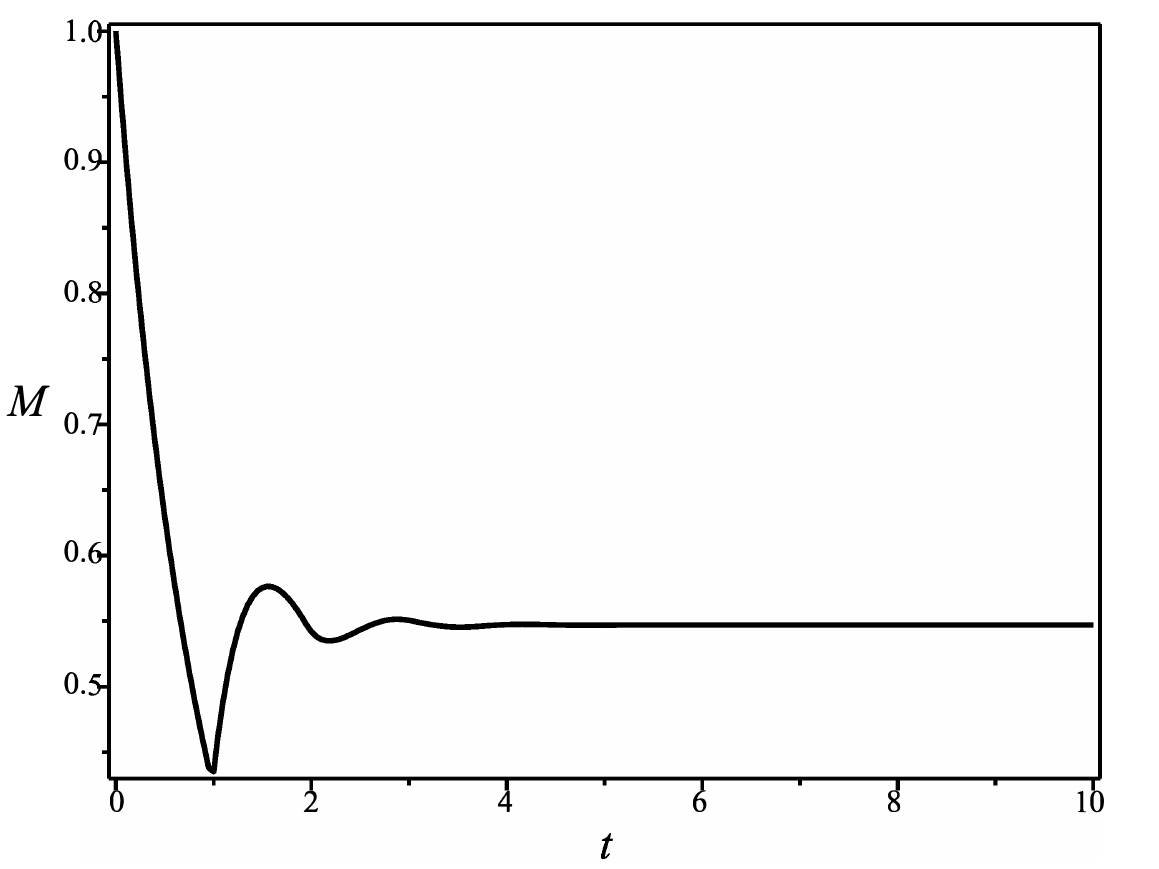}
  \includegraphics[scale=0.45]{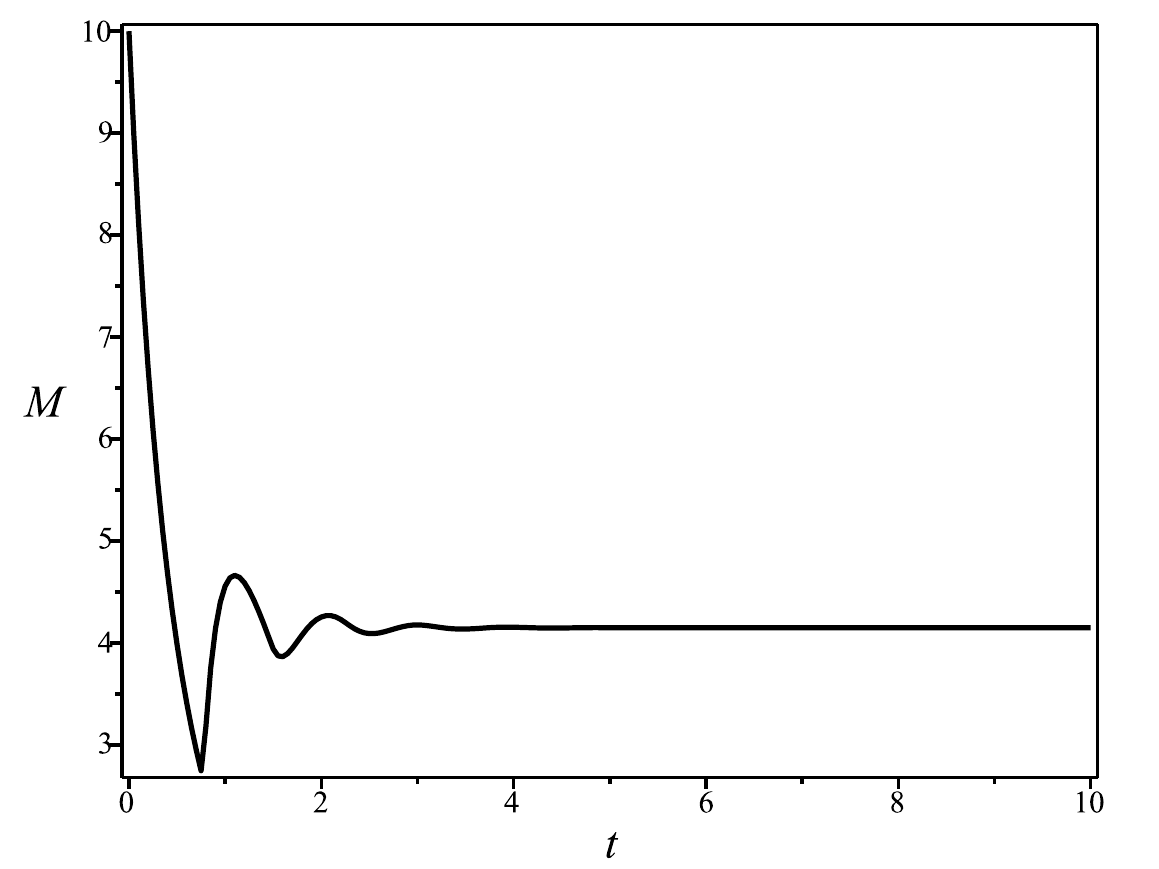}
  \includegraphics[scale=0.45]{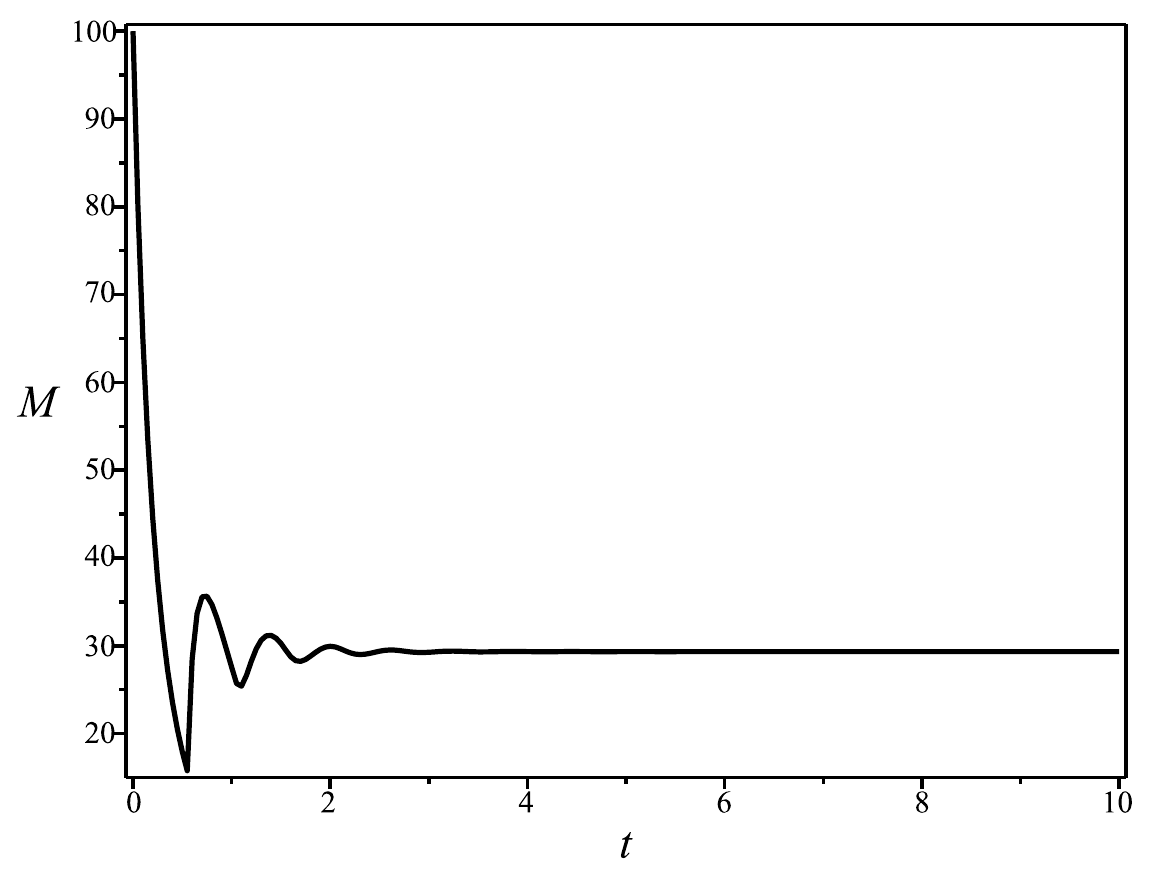}
  \caption{The black hole evolution at $\delta$=1. From left to right the initial mass of black hole are $M_\text{ini}= 1, 10, 100$ respectively.}\label{figurec}}
\end{figure}

\begin{figure}[thbp]
\center{
  \includegraphics[scale=0.45]{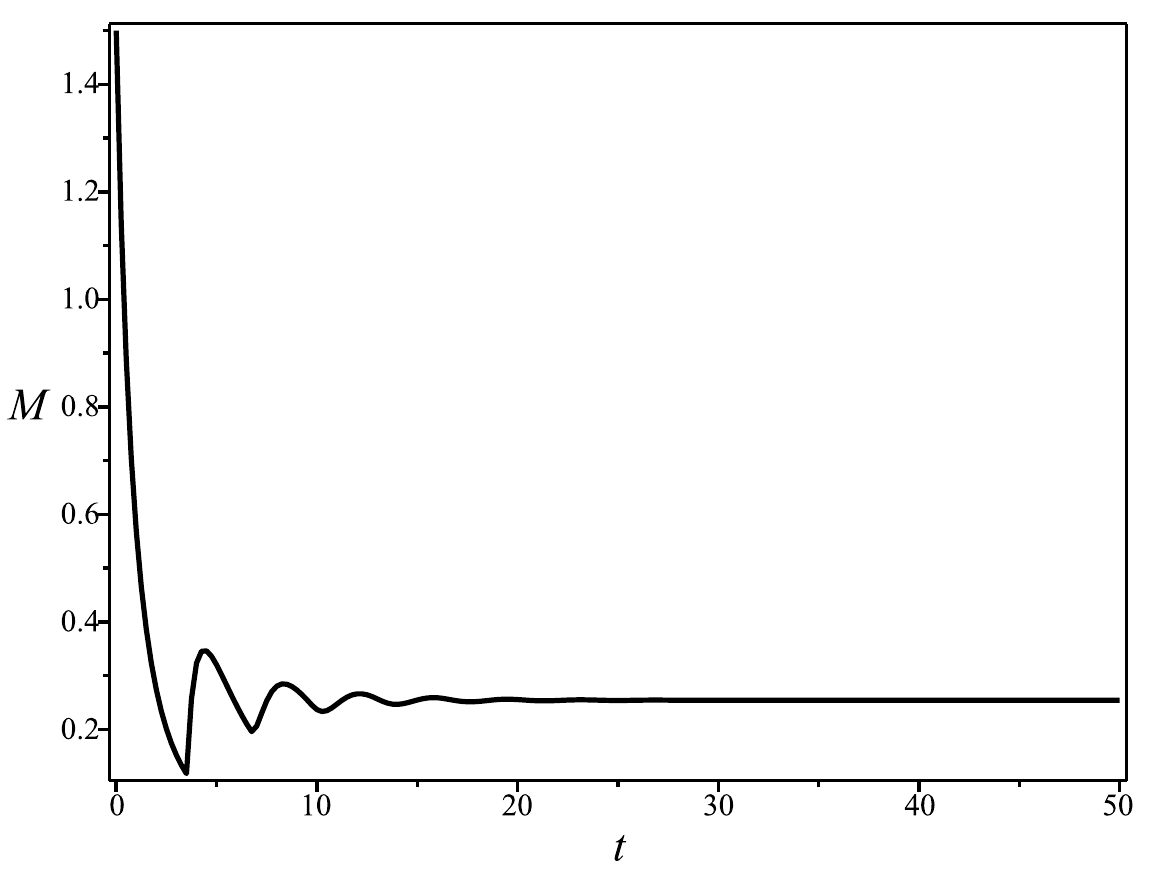}
  \includegraphics[scale=0.45]{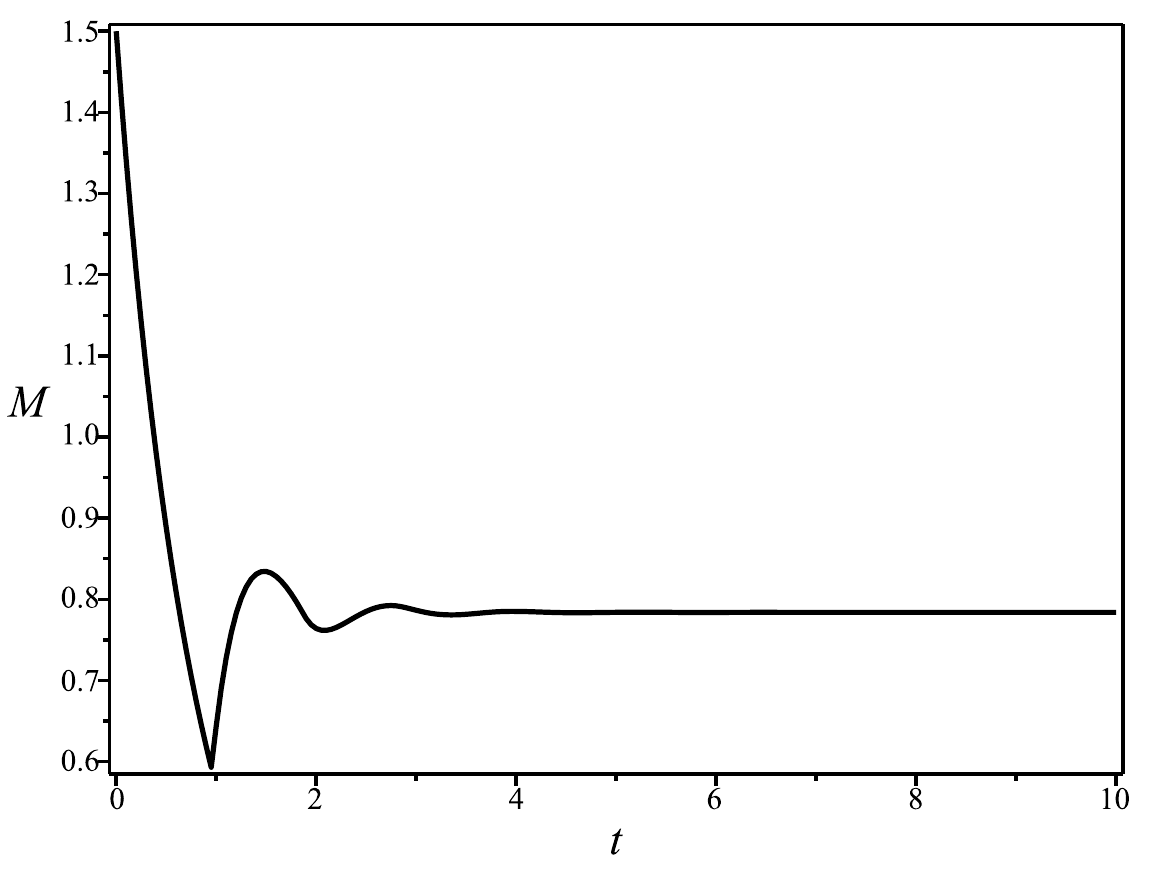}
  \includegraphics[scale=0.45]{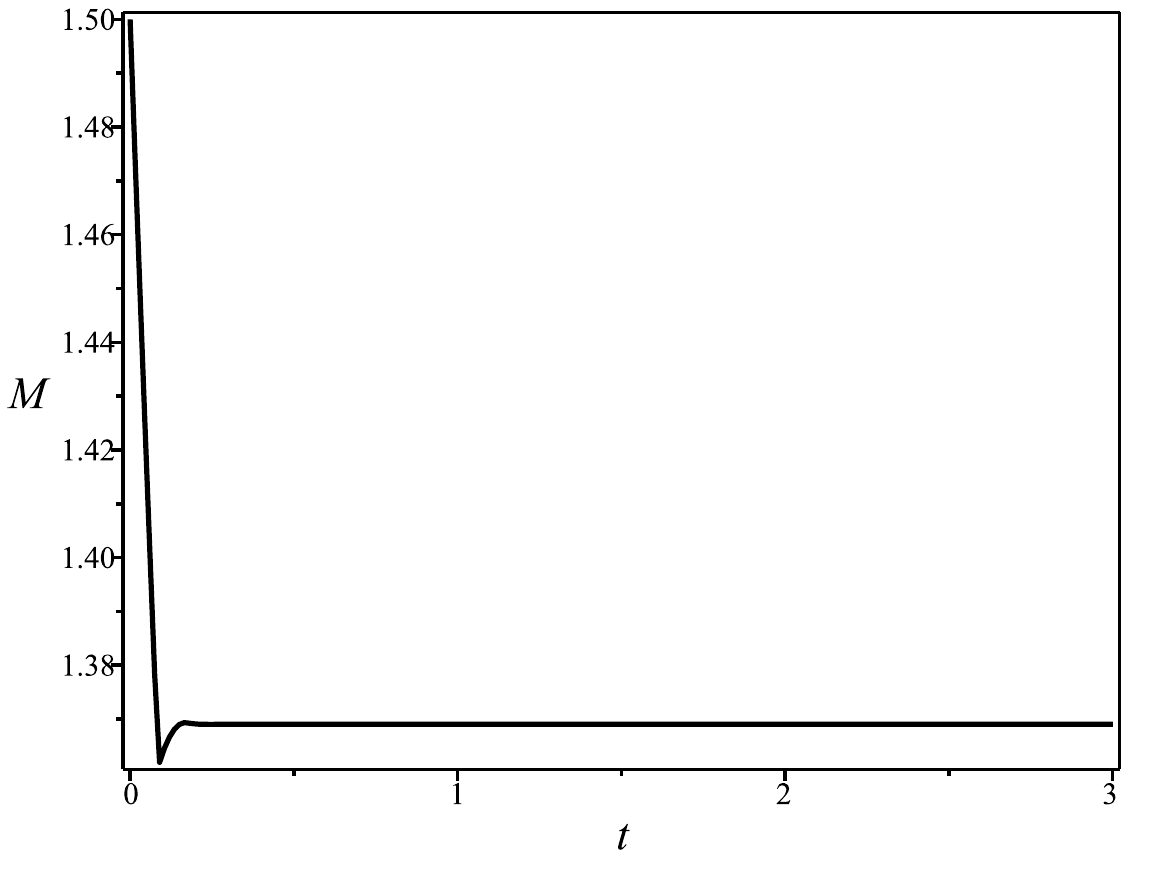}
  \caption{The black hole evolution at $M_\text{ini}$=1.5. From left to right the $\delta= 0.1, 1, 5$ respectively.}\label{figurecc}}
\end{figure}

Now we can solve the DDE numerically and study the black hole evolution. In Fig. (\ref{figurec}) and Fig. (\ref{figurecc}) we present some examples with different choices of the values of $\delta$ and $M_\text{ini}$. The results are consistent with Fig. (\ref{figureh}) and Fig. (\ref{figurehh}) in the case of fixed $t^*$. 

\begin{figure}[thbp]
\center{
 \includegraphics[scale=0.41]{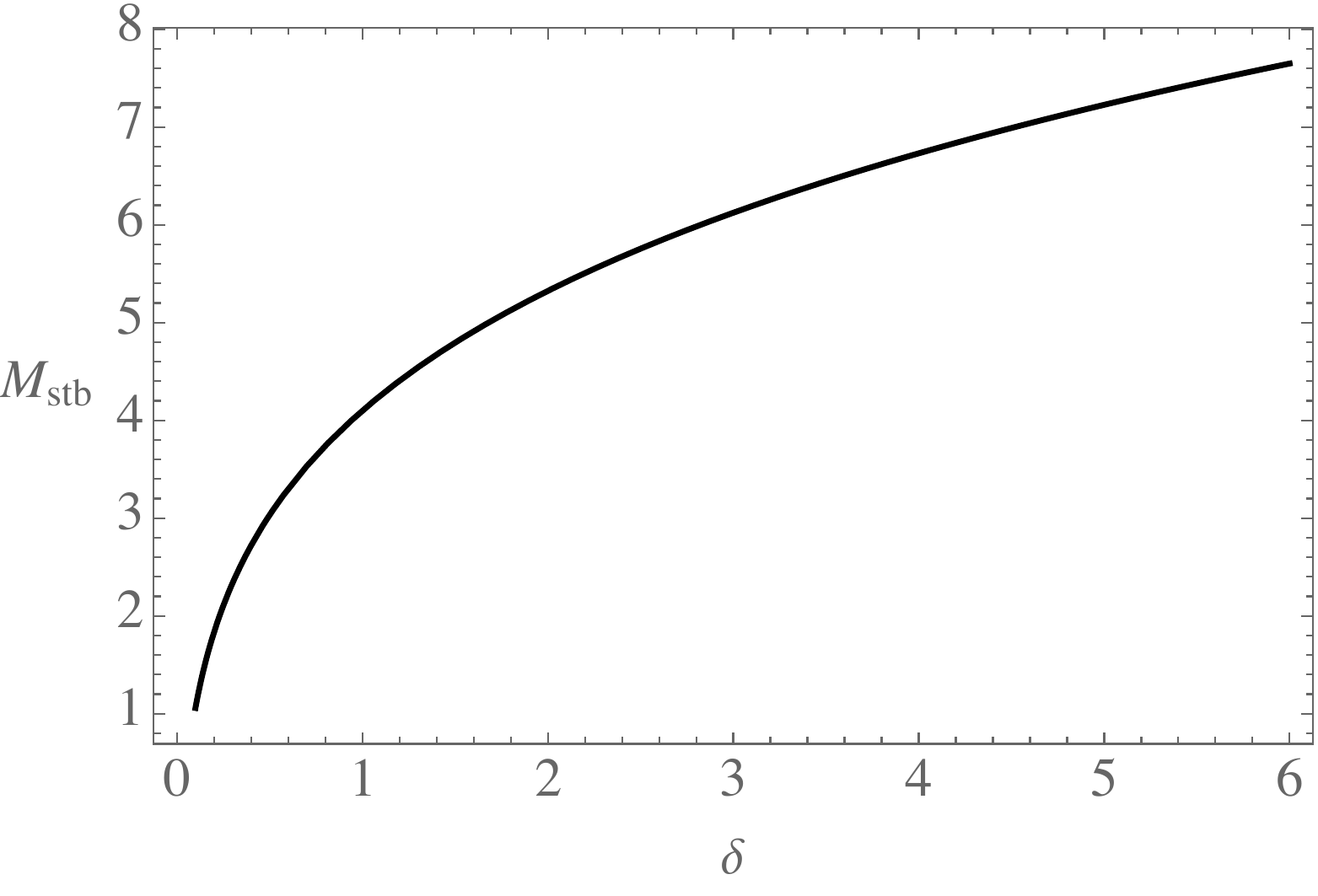}
  \includegraphics[scale=0.41]{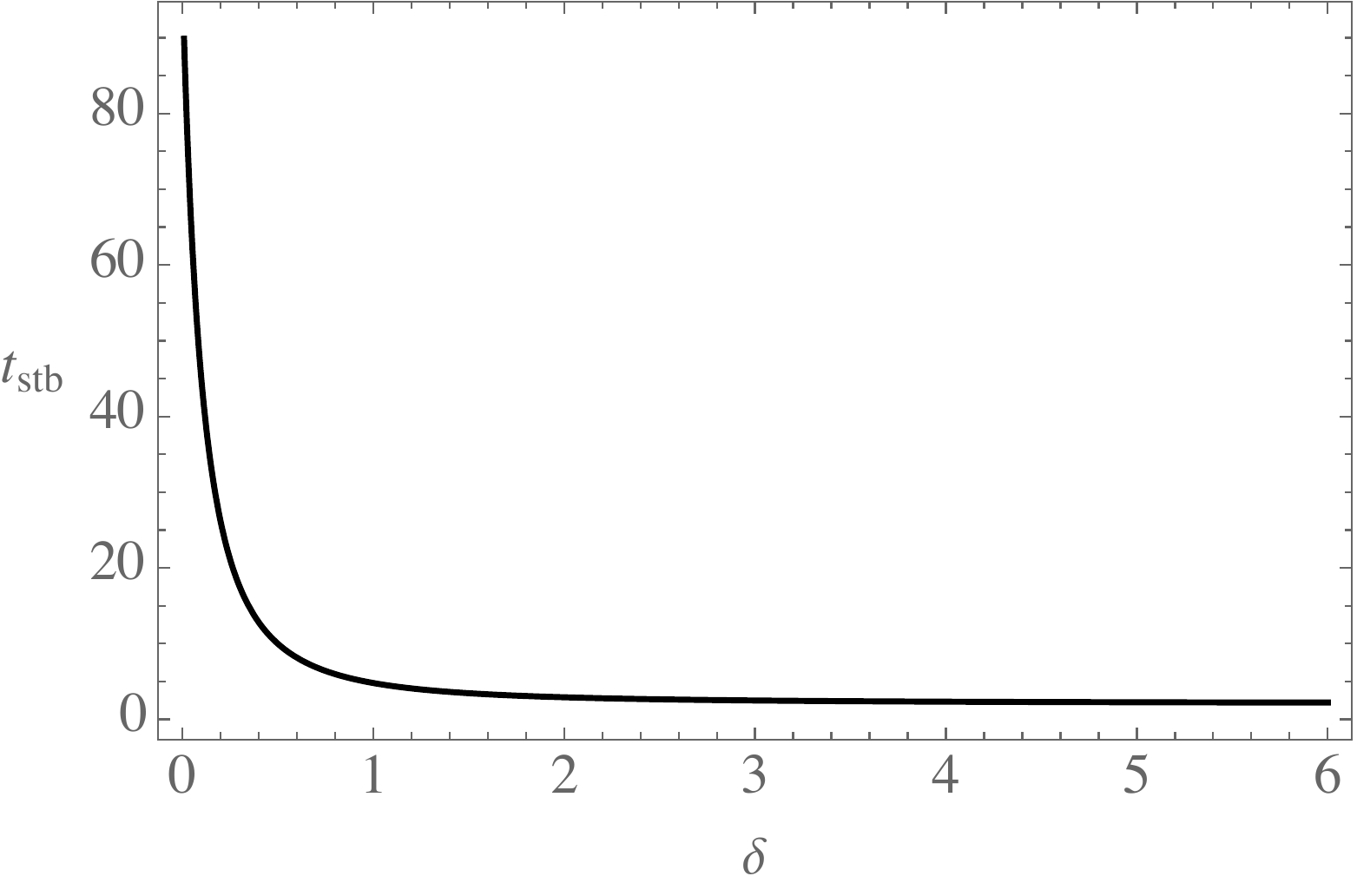}
  \caption{Left: The relationship between $\delta$ and the stable mass $M_\text{stb}$.; Right: The relationship between $\delta$ and the stable time $t_\text{stb}$. In both figures we set $M_\text{ini}=10$ and $L=1$.}\label{figured}}
\end{figure}

Similarly, we can analyze the behaviour of the final stable mass $M_\text{stb}$ and corresponding time $t_\text{stb}$ under the condition of varying $t^*$. In the left figure of Fig. (\ref{figured}), as $\delta$ increases, $M_\text{stb}$ also increases. This is because as $\delta$ grows, particles are emitted farther away from the event horizon, and so $t^*$ is  smaller, which in turn implies that it is harder for the black hole to evaporate.

The right side of Fig. (\ref{figured}) shows that $t_\text{stb}$ decreases with $\delta$. The black hole mass decreases rapidly when $\delta$ is small. This is because when $\delta\to 0$ , as we have explained, $t^*\to \infty$, thus the evolution is dominated by the first equation of the DDE in Eq. (\ref{DDE}) and the situation reduces to that of complete absorptive boundary condition. For small but nonzero $\delta$, the oscillation in the black hole  mass is more obvious, but eventually, it will reach a stable state, as shown in the left figure of Fig. (\ref{figurecc}). When $\delta$ is larger, the black hole oscillation is milder and it reaches a stable state sooner. This is consistent with the situation with fixed $t^*$ (Fig. (\ref{figuremm})) since $t^*$ is inversely proportional to $\delta$.

\begin{figure}[thbp]
\center{
  \includegraphics[scale=0.445]{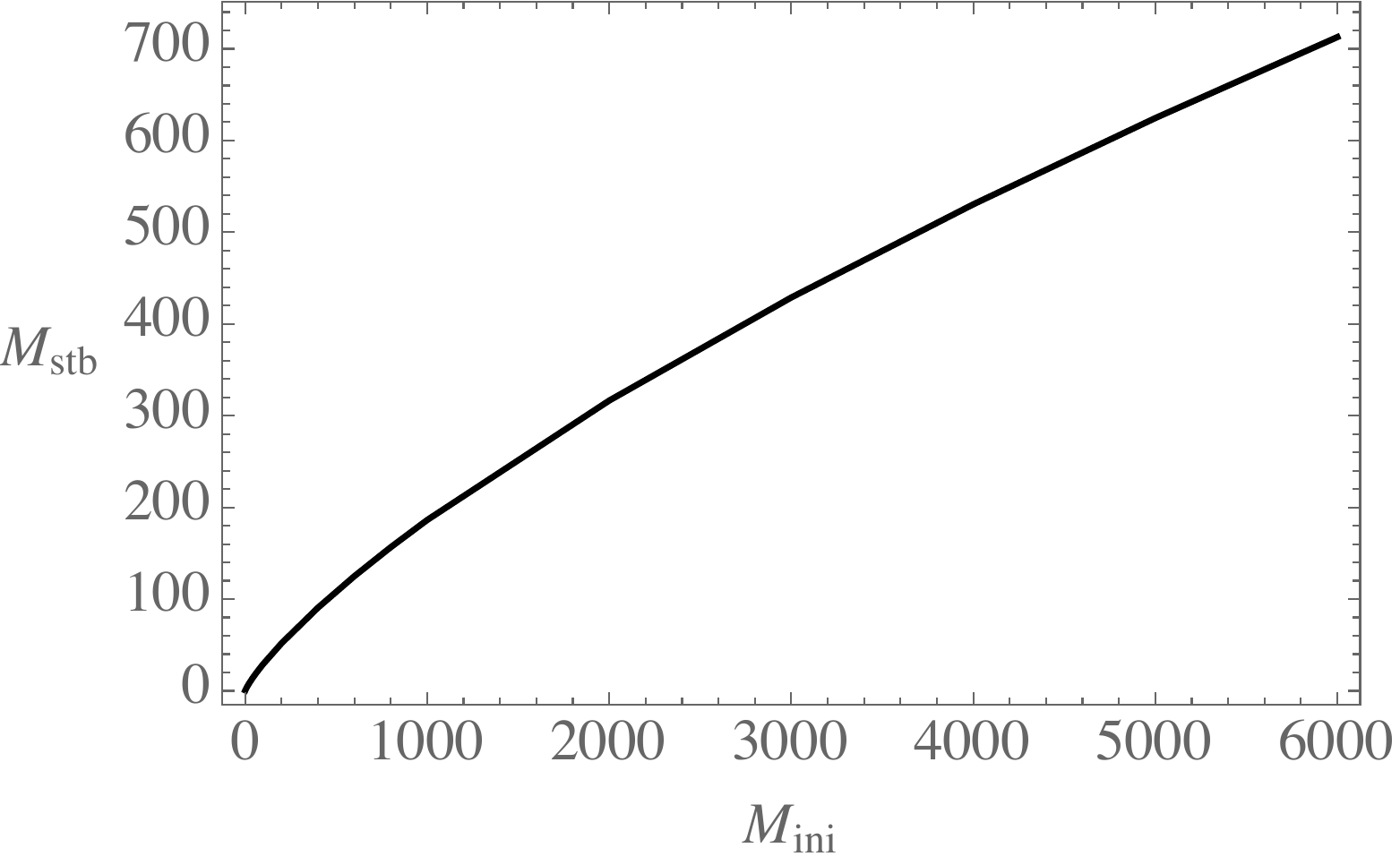}
  \includegraphics[scale=0.42]{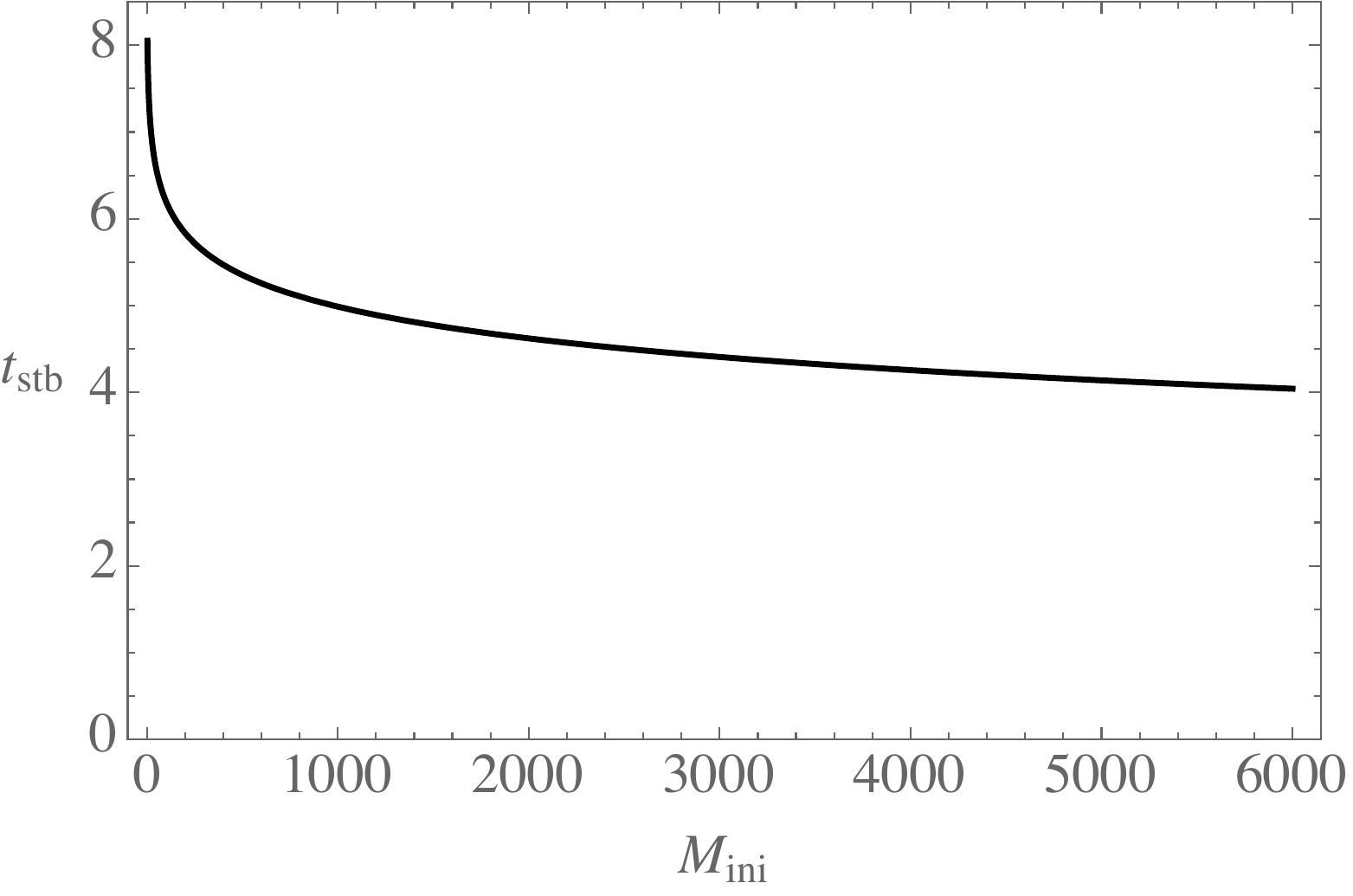}
  \caption{
The relationship between the initial mass $M_\text{ini}$ and final stable mass $M_\text{stb}$; Right: The relationship between the initial mass $M_\text{ini}$ and the final stable time $t_\text{stb}$. In both figures we set $\delta=1$ and $L=1$.
}\label{figuref}}
\end{figure}

In Fig. (\ref{figuref}), we present the behaviors of $M_\text{stb}$ and $t_\text{stb}$ with various initial black hole mass $M_\text{ini}$ for fixed $\delta$. In the left figure we find that the final stable mass increases with the initial mass, which is to be expected. Notice that in the right figure,  $t_\text{stb}$ decreases with the growth of $M_\text{ini}$. Indeed, according to Eq. (\ref{tstar}), the black hole mass $M$ appears in the denominator, thus a larger $M_\text{ini}$ predicts a smaller $t^*$, so the duration of the whole evaporation process will decrease, and the time to reach thermal equilibrium will also become shorter.

Note that the behaviors one obtains for the varying $t^*$ case are qualitatively very similar to the fixed $t^*$ case, except for the plot of $t_\text{stb}$ against $M_\text{ini}$, which shows a completely opposite behavior, c.f. the right plot of Fig. (\ref{figuref}) with the right plot of Fig. (\ref{figuremm}). We see in the right plot of Fig. (\ref{figuredd}) that the (fixed) $t^*$ should decrease as $t_\text{stb}$ decreases (for fixed $M_\text{ini}$), whereas in the right plot of Fig. (\ref{figuremm}), for fixed $t^*=1$, we observe that $t_\text{stb}$ increases with $M_\text{ini}$. Thus one must consider these two effects together carefully, and from the varying $t^*$ case we see that the overall effect is still for $t_\text{stb}$ to decrease with $M_\text{ini}$.

\section{Discussion}

Since asymptotically locally anti-de Sitter spacetimes have a timelike boundary, it is not globally hyperbolic. This means that one must prescribe \red{a} boundary condition on the conformal boundary in order to solve for the evolution of some given initial datum on a spacelike hypersurface. The usual choice is to impose a completely reflective boundary condition. Under such an assumption, the massless Hawking emission will reach the boundary and be reflected back into the black hole. As a result, large black holes are believed to be in equilibrium with their Hawking radiation. The study of how black holes in AdS can reach thermal equilibrium is also closely related to the echoes of gravitational waves in the bulk (and its possible holographic dual interpretation), which has recently gained some attention in the literature \cite{Saraswat:2019npa, 2007.01879}.

In this work, applying a completely reflective boundary condition, we modeled the Hawking evaporation of flat black holes in AdS with a simple DDE. Note that we assume the black hole is pre-existing and that Hawking radiation suddenly switches on at a particular time. The results could therefore be somewhat different if one considers instead a fully dynamical gravitational collapse that forms a black hole (in that context, there is also a debate concerning the importance of the so-called ``pre-Hawking radiation'' \cite{1610.07839,1710.01533,1802.09107}). This would require setting up a model with a Vaidya-like metric. We are also assuming the Stefan-Boltzmann equation holds for all time, and that the emitted radiation follows null geodesics, i.e. we are assuming the geometric optics approximation. Of course, the fine print is that such an approximation will eventually break down when the black hole is small enough. Nevertheless, as a first step to concretely model the evaporation, we work within the simplest model possible to gain some insights into the general picture\footnote{``All models are wrong, but some are useful.'' -- George E. P. Box.}. The model can be made more sophisticated if we take into account the changes of the horizon position and therefore the blueshift of the incoming radiation does not perfectly cancel with the redshift of the outgoing radiation. 

From our results, we see that the typical behavior is that the mass of the black hole displays a damped oscillation behavior as a function of the coordinate time $t$, and eventually asymptotes to a stable state corresponding to thermal equilibrium. Perhaps what is more surprising, is that a large black hole can evaporate into a small black hole that eventually reaches thermal equilibrium with its Hawking radiation, whereas the usual folklore seems to give the impression that a large black hole stays big. There is no contradiction however, since our analysis is restricted to black holes with toroidal topology. Small toroidal black holes are cold, not hot, so they have a chance to reach thermal equilibrium. 

Black holes with $k=1$ and $k=-1$ are expected to behave differently. For example, we know that for $k=1$, small black holes are hot, so they are expected to completely evaporate even under a completely reflective boundary condition. However, large AdS black holes are hot regardless of their topologies. Therefore we expect that at least some of the qualitative pictures we found here continue to hold. 
However, for the $k=1$ case we would expect that large black holes should be able to achieve thermal equilibrium (and stays large).
It would be interesting to verify this with our (very) simplified model. If we do \emph{not} obtain the expected result, this might indicate that the model is too simplistic. This is beyond our current work however, as the $k=\pm 1$ cases are difficult to solve with our current method (DDE solving with MAPLE and Mathematica still cannot handle these deceptively simple equations, so a more sophisticated numerical method would be required in future work.) Note that for the $k=-1$ case, as is evident from Eq. (\ref{temp}), there is a minimial radius at which the temperature vanishes. This would correspond to a negative mass black hole \cite{9808032}. We expect that this state cannot be reached (the third law of black hole thermodynamics).

We also have not considered other effects that could arise in toroidal black hole spacetimes. For example, for a sufficiently small black hole, if its temperature drops below a critical temperature
\begin{equation}
T_c = \frac{1}{2\pi K L},
\end{equation}
the black hole would undergo a phase transition into the Horowitz-Myers soliton \cite{9808079,0108170,0204081,0905.1180} (this is essentially a toroidal analogue for the well-known Hawking-Page phase transition \cite{HP}), despite the fact that there is no minimum temperature for such black holes. 

Lastly, it is important to emphasize that in this work we have only discussed how the black hole mass evolves under Hawking evaporation.
There are a lot of subtleties in the underlying physics that can be further explored (even in the context of our ``minimalistic model''). Notably, one could look at the semi-classical stress-energy tensor and compute the flux of the particle production more precisely, even considering back-scattering in the stress-tensor (see, e.g. \cite{FN}), or backreaction on black hole \cite{Susskind:1992gd}. 
It would be interesting also to see how the energy condition is explicitly broken and how we can observe this throughout the evaporation process. A related question would be to see if there is any negative energy flux coming out from the black hole especially when quantum information aspects are considered \cite{1404.0602,1405.5235,1506.08072}.
Indeed, in view of the recent trend to incorporate quantum information into the study of black hole physics (partly motivated by the black hole information paradox), an ambitious aim would be to track how various quantities like entanglement entropy, complexity, and mutual information evolve along with the black hole mass. Note that this would require us to focus on the entire system, namely both the black hole and the thermal bath, instead of just focusing on the mass loss and mass gain of the black hole.

\begin{acknowledgments}
YCO thanks the National Natural Science Foundation of China (No.11922508) for funding support.
He also thanks Brett McInnes for useful discussions.
Hao Xu thanks Natural Science Foundation of the Jiangsu Higher Education Institutions of China (Grant No.20KJD140001) for funding support. 
\end{acknowledgments}


\begin{thebibliography}{99}
\baselineskip=0.6 cm

\bibitem{9808032}
Danny Birmingham, ``Topological Black Holes in Anti-de Sitter Space'', {\hypersetup{urlcolor=vividviolet}\href{https://iopscience.iop.org/article/10.1088/0264-9381/16/4/009}{Class. Quant. Grav. \textbf{16} (1999) 1197}}, \href{https://arxiv.org/abs/hep-th/9808032}{[arXiv:hep-th/9808032]}.

\bibitem{0709.3738}
Samuli Hemming, Larus Thorlacius, ``Thermodynamics of Large AdS Black Holes'', {\hypersetup{urlcolor=vividviolet}\href{https://iopscience.iop.org/article/10.1088/1126-6708/2007/11/086}{JHEP \textbf{11} (2007) 086}}, \href{https://arxiv.org/abs/0709.3738}{[arXiv:0709.3738 [hep-th]]}.

\bibitem{0805.1876}
Erling J. Brynjolfsson, Larus Thorlacius, ``Taking the Temperature of a Black Hole'', {\hypersetup{urlcolor=vividviolet}\href{https://iopscience.iop.org/article/10.1088/1126-6708/2008/09/066}{JHEP \text{09} (2008) 066}}, \href{https://arxiv.org/abs/0805.1876}{[arXiv:0805.1876 [hep-th]]}.

\bibitem{0911.4144}
Veronika E Hubeny, Donald Marolf, Mukund Rangamani, ``Hawking Radiation From AdS Black Holes'', {\hypersetup{urlcolor=vividviolet}\href{https://iopscience.iop.org/article/10.1088/0264-9381/27/9/095018}{Class. Quant. Grav. \textbf{27} (2010) 095018}}, \href{https://arxiv.org/abs/0911.4144}{[arXiv:0911.4144 [hep-th]]}.

\bibitem{1104.3702}
Piotr Bizo\'n, Andrzej Rostworowski, ``Weakly Turbulent Instability of Anti–de Sitter Spacetime'', {\hypersetup{urlcolor=vividviolet}\href{https://journals.aps.org/prl/abstract/10.1103/PhysRevLett.107.031102}{Phys. Rev. Lett. \textbf{107} (2011) 031102}}, \href{https://arxiv.org/abs/1104.3702}{[arXiv:1104.3702 [gr-qc]]}.

\bibitem{1708.05600}
Gr\'egoire Martinon, ``The Instability of Anti-de Sitter Space-time'', \href{https://arxiv.org/abs/1708.05600}{[arXiv:1708.05600 [gr-qc]]}.

\bibitem{3343}
Gary W. Gibbons, Malcolm J. Perry, ``Black Holes in Thermal Equilibrium'', {\hypersetup{urlcolor=vividviolet}\href{https://journals.aps.org/prl/abstract/10.1103/PhysRevLett.36.985}{Phys. Rev. Lett. \textbf{36} (1976) 985}}.

\bibitem{1507.07845}
Yen Chin Ong, ``Hawking Evaporation Time Scale of Topological Black Holes in Anti-de Sitter Spacetime'', {\hypersetup{urlcolor=vividviolet}\href{https://www.sciencedirect.com/science/article/pii/S0550321316000067?via\%3Dihub}{Nucl. Phys. B \textbf{903} (2016) 387}}, \href{https://arxiv.org/abs/1507.07845}{[arXiv:1507.07845 [gr-qc]]}.

\bibitem{1304.6483}
Ahmed Almheiri, Donald Marolf, Joseph Polchinski, Douglas Stanford, James Sully, ``An Apologia for Firewalls'', {\hypersetup{urlcolor=vividviolet}\href{https://link.springer.com/article/10.1007/JHEP09(2013)018}{JHEP \textbf{1309} (2013) 018}}, \href{http://arxiv.org/abs/1304.6483}{[arXiv:1304.6483 [hep-th]]}.

\bibitem{0804.0055}
Jorge V. Rocha, ``Evaporation of Large Black Holes in AdS: Coupling to the Evaporon'', {\hypersetup{urlcolor=vividviolet}\href{https://iopscience.iop.org/article/10.1088/1126-6708/2008/08/075}{JHEP \textbf{0808} (2008) 075}}, \href{http://arxiv.org/abs/0804.0055}{[arXiv:0804.0055 [hep-th]]}.

\bibitem{1307.1796}
Mark Van Raamsdonk, ``Evaporating Firewalls'', {\hypersetup{urlcolor=vividviolet}\href{https://link.springer.com/article/10.1007\%2FJHEP11\%282014\%29038}{JHEP \textbf{11} (2014) 038}}, \href{http://arxiv.org/abs/1307.1796}{[arXiv:1307.1796 [hep-th]]}.

\bibitem{PhysRevD.18.3565}
S. J. Avis, C. J. Isham, D. Storey, {\hypersetup{urlcolor=vividviolet}\href{https://journals.aps.org/prd/abstract/10.1103/PhysRevD.18.3565}{Phys. Rev. D \textbf{18} (1978) 3565}}, ``Quantum Field Theory in Anti-de Sitter Space-Time''.

\bibitem{0402184}
Akihiro Ishibashi, Robert M. Wald, ``Dynamics in Non-Globally-Hyperbolic Static Spacetimes III: Anti-de Sitter Spacetime'', {\hypersetup{urlcolor=vividviolet}\href{https://iopscience.iop.org/article/10.1088/0264-9381/21/12/012}{Class. Quant. Grav. \textbf{21} (2004) 2981}}, \href{https://arxiv.org/abs/hep-th/0402184}{[arXiv:hep-th/0402184]}.

\bibitem{1302.1580}
Oscar J.C. Dias, Jorge E. Santos, ``Boundary Conditions for Kerr-AdS Perturbations'', {\hypersetup{urlcolor=vividviolet}\href{https://link.springer.com/article/10.1007/JHEP10(2013)156}{JHEP \textbf{10} (2013) 156}}, \href{https://arxiv.org/abs/1302.1580}{[arXiv:1302.1580 [hep-th]]}.

\bibitem{1701.01119}
M. C. Baldiotti, R. Fresneda, C. Molina, ``A Hamiltonian Approach for the Thermodynamics of Ads Black Holes'', {\hypersetup{urlcolor=vividviolet}\href{https://www.sciencedirect.com/science/article/abs/pii/S0003491617301215?via\%3Dihub}{Ann. Phys. \textbf{382} (2017) 22}}, \href{https://arxiv.org/abs/1701.01119}{[arXiv:1701.01119 [hep-th]]}.

\bibitem{9404041}
J. P. S. Lemos, ``Cylindrical black hole in general relativity'', {\hypersetup{urlcolor=vividviolet}\href{https://www.sciencedirect.com/science/article/abs/pii/037026939500533Q}{Phys. Lett. B \textbf{353} (1995), 46-51}}, \href{https://arxiv.org/abs/gr-qc/9404041}{[arXiv:gr-qc/9404041]}.

\bibitem{Huang:1995zb}
C. G. Huang and C. B. Liang, ``A Torus like black hole'', {\hypersetup{urlcolor=vividviolet}\href{https://www.sciencedirect.com/science/article/abs/pii/037596019500229V}{Phys. Lett. A \textbf{201} (1995), 27-32}}.

\bibitem{9609065}
R. G. Cai and Y. Z. Zhang, ``Black plane solutions in four-dimensional space-times'', {\hypersetup{urlcolor=vividviolet}\href{https://journals.aps.org/prd/abstract/10.1103/PhysRevD.54.4891}{Phys. Rev. D \textbf{54} (1996), 4891-4898}}, \href{https://arxiv.org/abs/gr-qc/9609065}{[arXiv:gr-qc/9609065]}.

\bibitem{1403.4886}
Yen Chin Ong, Brett McInnes, Pisin Chen, ``Cold Black Holes in the Harlow-Hayden Approach to Firewalls'', {\hypersetup{urlcolor=vividviolet}\href{https://linkinghub.elsevier.com/retrieve/pii/S0550321314004027}{Nucl. Phys. B \textbf{891} (2015) 627}}, \href{http://arxiv.org/abs/1403.4886}{[arXiv:1403.4886 [hep-th]]}.

\bibitem{9803061}
Dietmar Klemm, Luciano Vanzo, ``Quantum Properties of Topological Black Holes'', {\hypersetup{urlcolor=vividviolet}\href{https://journals.aps.org/prd/abstract/10.1103/PhysRevD.58.104025}{Phys. Rev. D \textbf{58} (1998) 104025}}, \href{http://arxiv.org/abs/gr-qc/9803061}{[arXiv:gr-qc/9803061]}.

\bibitem{1507.02682}
Don N. Page, ``Finite Upper Bound for the Hawking Decay Time of an Arbitrarily Large Black Hole in Anti-de Sitter Spacetime'', {\hypersetup{urlcolor=vividviolet}\href{https://journals.aps.org/prd/abstract/10.1103/PhysRevD.97.024004}{Phys. Rev. D \textbf{97} (2018) 024004}}, \href{https://arxiv.org/abs/1507.02682}{[arXiv:1507.02682 [hep-th]]}.

\bibitem{Saraswat:2019npa}
  Krishan Saraswat, Niayesh Afshordi,
  ``Quantum Nature of Black Holes: Fast Scrambling versus Echoes'', 
  {\hypersetup{urlcolor=vividviolet}\href{https://link.springer.com/article/10.1007\%2FJHEP04\%282020\%29136}{JHEP {\bf 2004} (2020) 136}},
  \href{https://arxiv.org/abs/1906.02653}{[arXiv:1906.02653 [hep-th]]}.


\bibitem{2003.10429}
Yen Chin Ong, Michael R. R. Good, ``The Quantum Atmosphere of Reissner-Nordstr\"om Black Holes'', 	{\hypersetup{urlcolor=vividviolet}\href{https://journals.aps.org/prresearch/abstract/10.1103/PhysRevResearch.2.033322}{Phys. Rev. Research \textbf{2}  (2020) 033322}}, \href{https://arxiv.org/abs/2003.10429}{[arXiv:2003.10429 [gr-qc]]}.

\bibitem{1511.08221}
Steven B. Giddings, ``Hawking Radiation, the Stefan-Boltzmann Law, and Unitarization'', {\hypersetup{urlcolor=vividviolet}\href{https://www.sciencedirect.com/science/article/pii/S0370269316000022?via\%3Dihub}{Phys. Lett. B \textbf{754} (2016) 39}}, \href{https://arxiv.org/abs/1511.08221v3}{[	arXiv:1511.08221 [hep-th]]}.

\bibitem{1607.02510}
Shahar Hod, ``Hawking Radiation and the Stefan-Boltzmann Law: The Effective Radius of the Black-Hole Quantum Atmosphere'', {\hypersetup{urlcolor=vividviolet}\href{https://www.sciencedirect.com/science/article/pii/S0370269316300429?via\%3Dihub}{Phys. Lett. B \textbf{757} (2016) 121}}, \href{https://arxiv.org/abs/1607.02510}{[arXiv:1607.02510 [gr-qc]]}.

\bibitem{1701.06161}
Ramit Dey, Stefano Liberati, Daniele Pranzetti, ``The Black Hole Quantum Atmosphere'', {\hypersetup{urlcolor=vividviolet}\href{https://linkinghub.elsevier.com/retrieve/pii/S0370269317307888}{Phys .Lett. B \textbf{774} (2017) 308}}, \href{https://arxiv.org/abs/1701.06161}{[arXiv:1701.06161 [gr-qc]]}.


\bibitem{2007.01879}
Vasil Dimitrov, Tom Lemmens, Daniel R. Mayerson, Vincent S. Min, Bert Vercnocke, ``Gravitational Waves, Holography, and Black Hole Microstates'', \href{https://arxiv.org/abs/2007.01879}{[arXiv:2007.01879 [hep-th]]}.

\bibitem{1610.07839}
Valentina Baccetti, Robert B. Mann, Daniel R. Terno, ``Role of Evaporation in Gravitational Collapse'', {\hypersetup{urlcolor=vividviolet}\href{https://iopscience.iop.org/article/10.1088/1361-6382/aad70e}{Class. Quantum Grav. \textbf{35} (2018) 185005}}, \href{https://arxiv.org/abs/1610.07839}{[arXiv:1610.07839 [gr-qc]]}.

\bibitem{1710.01533}
Pisin Chen, William G. Unruh, Chih-Hung Wu, Dong-han Yeom, ``Pre-Hawking Radiation Cannot Prevent the Formation of Apparent Horizon'', {\hypersetup{urlcolor=vividviolet}\href{https://journals.aps.org/prd/abstract/10.1103/PhysRevD.97.064045}{Phys. Rev. D 97 (2018) 064045}}, \href{https://arxiv.org/abs/1710.01533}{[arXiv:1710.01533 [gr-qc]]}.

\bibitem{1802.09107}
William G Unruh, ``Prehawking Radiation'', \href{https://arxiv.org/abs/1802.09107}{[arXiv:1802.09107 [gr-qc]]}.

\bibitem{9808079}
Gary T. Horowitz, Robert C. Myers, ``The AdS/CFT Correspondence and a New Positive Energy Conjecture for General Relativity'', {\hypersetup{urlcolor=vividviolet}\href{https://journals.aps.org/prd/abstract/10.1103/PhysRevD.59.026005}{Phys. Rev. D \textbf{59} (1998) 026005}}, \href{https://arxiv.org/abs/hep-th/9808079}{[arXiv:hep-th/9808079]}.

\bibitem{0108170}
Gregory J. Galloway, Eric Woolgar, Sumati Surya, ``A Uniqueness Theorem for the AdS Soliton'', {\hypersetup{urlcolor=vividviolet}\href{https://journals.aps.org/prl/abstract/10.1103/PhysRevLett.88.101102}{Phys. Rev. Lett. \textbf{88} (2002) 101102}}, \href{https://arxiv.org/abs/hep-th/0108170}{[arXiv:hep-th/0108170]}.

\bibitem{0204081}
Gregory J. Galloway, Sumati Surya, Eric Woolgar, ``On the Geometry and Mass of Static, Asymptotically AdS Spacetimes, and the Uniqueness of the AdS Soliton'', {\hypersetup{urlcolor=vividviolet}\href{https://link.springer.com/article/10.1007\%2Fs00220-003-0912-7}{Commun. Math. Phys. \textbf{241} (2003) 1}}, \href{https://arxiv.org/abs/hep-th/0204081}{[arXiv:hep-th/0204081]}.

\bibitem{0905.1180}
Brett McInnes, ``Bounding the Temperatures of Black Holes Dual to Strongly Coupled Field Theories on Flat Spacetime'', {\hypersetup{urlcolor=vividviolet}\href{http://iopscience.iop.org/1126-6708/2009/09/048/pdf/jhep092009048.pdf}{JHEP \textbf{09} (2009) 048}}, \href{https://arxiv.org/abs/0905.1180}{[arXiv:0905.1180 [hep-th]]}.

\bibitem{HP}
Stephen W. Hawking, Don N. Page, ``Thermodynamics of Black Holes in Anti-de Sitter Space'', {\hypersetup{urlcolor=vividviolet}\href{https://link.springer.com/article/10.1007/BF01208266}{Commun. Math. Phys. \textbf{87} (1983) 577}}.

\bibitem{FN}
Alessandro Fabbri, Jos\'e Navarro-Salas, \emph{Modeling Black Hole Evaporation}, Imperial College Press, 2005.


\bibitem{Susskind:1992gd}
Leonard Susskind, L\'arus Thorlacius, ``Hawking Radiation and Back-Reaction'',
{\hypersetup{urlcolor=vividviolet}\href{https://doi.org/10.1016/0550-3213(92)90081-L}{Nucl. Phys. B \textbf{382} (1992) 123}}, \href{https://arxiv.org/abs/hep-th/9203054}{[arXiv:hep-th/9203054 [hep-th]]}.

\bibitem{1404.0602}
Eugenio Bianchi, Matteo Smerlak, ``Entanglement Entropy and Negative Energy in Two Dimensions'', {\hypersetup{urlcolor=vividviolet}\href{https://journals.aps.org/prd/abstract/10.1103/PhysRevD.90.041904}{Phys. Rev. D \textbf{90} (2014) 041904}}, \href{https://arxiv.org/abs/1404.0602}{[arXiv:1404.0602 [gr-qc]]}.

\bibitem{1405.5235}
Eugenio Bianchi, Matteo Smerlak, ``Last Gasp of a Black Hole: Unitary Evaporation Implies Non-Monotonic Mass Loss'', 	{\hypersetup{urlcolor=vividviolet}\href{https://link.springer.com/article/10.1007\%2Fs10714-014-1809-9}{Gen. Relativ. Gravit. \textbf{46} (2014) 1809}}, \href{https://arxiv.org/abs/1405.5235}{[arXiv:1405.5235 [gr-qc]]}.


\bibitem{1506.08072}
Michael R.R. Good, Yen Chin Ong, ``Signatures of Energy Flux in Particle Production: A Black Hole Birth Cry and Death Gasp'', {\hypersetup{urlcolor=vividviolet}\href{https://link.springer.com/article/10.1007\%2FJHEP07\%282015\%29145}{JHEP \textbf{07} (2015) 145}}, \href{https://arxiv.org/abs/1506.08072}{[arXiv:1506.08072 [gr-qc]]}.

\end{thebibliography}
\end{document}